\documentclass[runningheads]{svjour3}                     % onecolumn (standard format)
\smartqed  % flush right qed marks, e.g. at end of proof
\usepackage{graphicx}
\usepackage[latin1]{inputenc}
\usepackage{color, colortbl}
\usepackage{natbib}
\usepackage[dvipdfm, pdfborder=0 0 0]{hyperref}
\usepackage{tikz}
\usepackage{algorithm,algorithmic}
\algsetup{indent=2em}

\usepackage{subfigure}
\DeclareGraphicsExtensions{.eps}

\newcommand\added[1]{\marginpar{\raggedright\footnotesize\sf\textcolor{blue}{\textbf{Added}}}\textcolor{blue}{#1}}
\newcommand\removed[1]{\marginpar{\raggedright\footnotesize\sf\textcolor{red}{\textbf{Removed}}}\textcolor{red!50}{#1}}
\definecolor{OliveGreen}{cmyk}{0.64,0,0.95,0.40}

\renewcommand\added[1]{#1}
\renewcommand\removed[1]{\strut}

\usepackage{amsmath}
\usepackage{amssymb}
\usepackage{amsfonts}
\usepackage{amsbsy}
\usepackage{bbm}

\usepackage[OT1]{fontenc}
\usepackage[charter]{mathdesign}
\usepackage[OMLmathbf,sfdefault=cmbr]{isomath}

\renewcommand{\eqref}[1]{Eq.~(\ref{#1})}

\newcommand{\cc}{{\mathcal C}}
\newcommand{\cd}{{\mathcal D}}

\newcommand{\cg}{{\mathcal G}}

\newcommand{\cl}{{\mathcal L}}
\newcommand{\cm}{{\mathcal M}}
\newcommand{\cn}{{\mathcal N}}

\newcommand{\cp}{{\mathcal P}}

\newcommand{\cx}{{\mathcal X}}
\newcommand{\cy}{{\mathcal Y}}

\newcommand{\Rr}{{\mathbb R}}

\newcommand{\ldeu}[2]{\cl_2\left(#1,\,#2\right)}

\newcommand{\di}[1]{{\rm d}#1} 				% d droit pour les integrandes
\newcommand{\ve}[1]{\boldsymbol{#1}}			% Vecteur
\newcommand{\ma}[1]{\boldsymbol{\rm #1}}		% Vecteur

\newcommand{\tr}{^{\textsf T}}				% Transpose

\newcommand{\enu}{ , \, \dots \,,}

\newcommand{\acc}[1]{\left\{#1\right\}}			% entre accolades {}
\newcommand{\Esp}[1]{{\mathbb E}\left[ #1 \right]}
\newcommand{\Espe}[2]{{\mathbb E}_{#1}\left[#2\right]}

\newcommand{\Pro}{\mathbb{P}}				% probabilité
\newcommand{\Prob}[1]{{\mathbb P}\left( #1 \right)}	% probabilité de ()
\newcommand{\fcar}[2] {{\mathbbm{1}}_{#1}\left(#2\right)}

\newcommand{\ie}{{\em i.e.} }
\newcommand{\eg}{{\em e.g.} }

\usepackage{multirow}                       % Cellules multilignes
\makeatletter
\def\hlinewd#1{%                                % Macro epaisseurs de lignes horizontales
\noalign{\ifnum0=`}\fi\hrule \@height #1 %
\futurelet\reserved@a\@xhline}
\makeatother
\newcommand{\hlineT}{\hlinewd{1.1pt}}
\newcommand{\hlineB}{\hlinewd{0.85pt}}

\journalname{Struct Multidisc Optim}

\begin{document}

\title{Reliability-based design optimization\\using kriging surrogates and subset simulation}
\titlerunning{RBDO using kriging surrogates and subset simulation}        % if too long for running head

\author{V. Dubourg \and B. Sudret \and J.-M. Bourinet}

\institute{V. Dubourg \and B.Sudret \at
              Phimeca Engineering, Centre d'Affaires du Zenith, 34 rue de Sarlieve, F-63800 Cournon d'Auvergne\\
	      Clermont Universite, IFMA, EA 3867, Laboratoire de Mecanique et Ingenieries, BP 10448, F-63000 Clermont-Ferrand\\
              \email{dubourg@phimeca.com, sudret@phimeca.com}
           \and
           J.-M. Bourinet \at
              Clermont Universite, IFMA, EA 3867, Laboratoire de Mecanique et Ingenieries, BP 10448, F-63000 Clermont-Ferrand\\
              \email{bourinet@ifma.fr}
}

\date{Received: 1 September 2010 / Revised: 18 February 2011 / Accepted: 30 March 2011.\\\textbf{Preprint submitted to Springer-Verlag.}}
% The correct dates will be entered by the editor

\maketitle

\begin{abstract}
The aim of the present paper is to develop a strategy for solving reliability-based design optimization (RBDO) problems that remains applicable when the performance models are expensive to evaluate\removed{(\eg a nonlinear finite element model used for the stability analysis of imperfect shells)}. Starting with the premise that simulation-based approaches are not affordable for such problems, and that the most-probable-failure-point-based approaches do not permit to quantify the error on the estimation of the failure probability, an approach based on both metamodels and advanced simulation techniques is explored. The kriging metamodeling technique is chosen in order to surrogate the performance functions because it allows one to genuinely quantify the surrogate error. The surrogate error onto the limit-state surfaces is propagated to the failure probabilities estimates in order to provide an empirical error measure. This error is then sequentially reduced by means of a population-based adaptive refinement technique until the kriging surrogates are accurate enough for reliability analysis. \added{This original refinement strategy makes it possible to add several observations in the design of experiments at the same time.} Reliability and reliability sensitivity analyses are performed by means of the \emph{subset simulation} technique for the sake of numerical efficiency. The adaptive surrogate-based strategy for reliability estimation is finally involved into a classical gradient-based optimization algorithm in order to solve the RBDO problem. The kriging surrogates are built in a so-called augmented reliability space thus making them reusable from one nested RBDO iteration to the other. The strategy is compared to other approaches available in the literature on three academic examples in the field of structural mechanics.
\keywords{reliability-based design optimization (RBDO) \and kriging \and surrogate modeling \and subset simulation \and adaptive refinement}
\end{abstract}

\section{Introduction}

In structural mechanics, design optimization is the decision-making process that aims at finding the best set of design variables which minimizes some cost model while satisfying some performance requirements. Due to the inconsistency between these two objectives, the optimal solutions often lie on the boundaries of the admissible space. Thus, these solutions are rather sensitive to uncertainty either in the parameters (\emph{aleatory}) or in the models themselves (\emph{epistemic}). \emph{Reliability-based design optimization} (RBDO) is a concept that accounts for uncertainty all along the optimization process. Basically, the deterministic performance model is wrapped into a more realistic probabilistic constraint which is referred to as the \emph{failure probability}. Despite its attractive formulation, the application field of RBDO is still limited to academic examples. This is mostly due to the fact that it is either based on simplifying assumptions that might not hold in practice; or in contrast, it requires computationally intensive stochastic simulations that are not affordable for real industrial problems. The present work attempts to propose an efficient strategy that would \emph{in fine} bring the RBDO application field to more sophisticated examples, closer to real engineering cases. In other words, the challenge is \emph{(i)} to provide an optimal safe design within a few hundred evaluations of the performance models and \emph{(ii)} to be able to quantify and minimize the errors induced by the various assumptions that are made along the development of the resolution strategy.\par

The remaining part of this introduction is devoted to the formulation of the RBDO problem the authors attempt to solve. A short literature review is also provided as an argument for the presently proposed surrogate-based resolution strategy. Section~\ref{sec:Kriging} introduces the kriging surrogate model. A specific emphasis is put on the \emph{epistemic nature} of the prediction error that is then used in Section~\ref{sec:DOE} in order to \emph{quantify} and \emph{minimize} the surrogate error. Section~\ref{sec:HSBRBDO} involves the adaptive refinement strategy of the kriging surrogate into a nested reliability-based design optimization loop. The convergence of the approach is finally heuristically demonstrated in Section~\ref{sec:Appli} through a few academic examples from the RBDO literature.\par

\subsection{Problem formulation}

Given a parametric model for the random vector $\ve{X}$ describing the environment of the system to be designed, the most basic formulation for the RBDO problem reads as follows:%
\begin{equation}\label{eq:RBDO_RIA}
 \ve{\theta}^* = \arg \min\limits_{\ve{\theta}\in\cd_{\ve{\theta}}} c\left(\ve{\theta}\right)\;:\quad
 \left\{\begin{array}{l}
   f_i\left(\ve{\theta}\right) \leq 0,\;i = 1\enu n_c \\
   \Prob{g_l\left(\ve{X}\left(\ve{\theta}\right)\right) \leq 0} \leq P_{f\,l}^0,\;l = 1\enu n_p
 \end{array}\right..
\end{equation}%
In this formulation, $c$ is the objective function to be minimized with respect to the design variables $\ve{\theta} \in \cd_{\ve{\theta}}$, while satisfying to $n_c$ deterministic soft constraints $\{f_i,\;i = 1\enu n_c\}$ bounding the so-called \emph{admissible design space} defined by the analyst. Note that in most applications these soft constraints consist in simple analytical functions that prevent the optimization algorithm from exploring regions of the design space that have no physical meaning (\eg negative or infinite dimensions), so that these constraints are inexpensive to evaluate. A \emph{deterministic design optimization} (DDO) problem would simply require additional performance functions $\{g_l,\;l = 1\enu n_p\}$ describing system failure with respect to the specific code of practice. As opposed to the previous soft constraints, these functions often involve the output of an expensive-to-evaluate black-box function $\cm$ -- \eg a finite element model. RBDO differs from DDO in the sense that these constraints are wrapped into $n_p$ probabilistic constraints $\{\Prob{g_l\left(\ve{X}(\ve{\theta})\right) \leq 0} \leq P_{f\,l}^0,\;l = 1\enu n_p\}$. $P_{f\,l}^0$ is the minimum safety requirement expressed here in the form of an acceptable \emph{probability of failure} which may be different for each performance function $g_l$. Such probabilities of failure are conveniently defined in terms of the following multidimensional integrals:
\begin{equation}\label{eq:pfdef}
  P_{f\,l}\left(\ve{\theta}\right) = \Prob{g_l\left(\ve{X}\left(\ve{\theta}\right)\right) \leq 0} = \int_{g_l\left(\ve{x}\right) \leq 0} f_{\ve{X}}\left(\ve{x},\,\ve{\theta}\right)\,\di{\ve{x}}, \quad l = 1\enu n_p.
\end{equation}\par

One should notice that, in the present formulation, the design vector $\ve{\theta}$ is a set of \emph{hyperparameters} defining the random vector $\ve{X}$. In other words, in this work, design variables are exclusively considered as hyperparameters in the joint probability density function $f_{\ve{X}}$ of the random vector $\ve{X}$ because it will later simplify the computation of \emph{reliability sensitivity} -- \ie the gradients of the failure probability. There is however no loss of generality since deterministic design variables might possibly be considered as artificially random (either normal or uniform) with small variance -- \ie sufficiently close to zero.\par

One could possibly argue that this formulation lacks full probabilistic consideration because the cost function is defined in a deterministic manner as it only depends on the hyperparameters $\ve{\theta}$ of the random vector $\ve{X}$. A more realistic formulation should eventually account for the randomness of the cost function possibly induced by the one in $\ve{X}$. However, the present formulation is extensively used in the RBDO literature for simplicity. Note however that thanks to the rather low complexity of usual cost models (analytical functions), an accurate simulation-based estimation of a mean cost, say $c\left(\ve{\theta}\right) = \Esp{c\left(\ve{X}\left(\ve{\theta}\right)\right)}$ would not require a large computational effort.\par

\subsection{Short literature review}

The most straightforward approach to solve the RBDO problem in \eqref{eq:RBDO_RIA} consists in nesting a reliability analysis within a nonlinear constrained optimization loop. Such methods are referred to as \emph{double-loop} or \emph{nested} approaches. Despite their conceptual simplicity these approaches are often argued to lack efficiency since they require too many evaluations of the performance functions $g_i$ that might involve the output of a time-consuming computer code. However for a broad range of applications where the performance functions are linear or weakly nonlinear -- \ie when the \removed{First Order Reliability Method} \added{\emph{first order reliability method} (FORM)} is applicable, the nested approach is able to give results within a reasonable number of evaluations of the performance functions (\eg \citet{Enevoldsen1994} achieve convergence within a few thousands evaluations).\par

The formulation in \eqref{eq:RBDO_RIA} is known as the \emph{reliability index approach} (RIA). Note that RIA often refers to nested RBDO algorithm based on FORM despite the fact that the concept can easily be extended to simulation-based reliability methods. Some authors, starting with \citet{Tu1999}, proposed an alternative formulation to the RBDO problem known as the \emph{performance measure approach} (PMA). The probabilistic constraints are transformed into quantile constraints that are approximated using the first-order reliability theory. According to \citet{Youn2004a}, PMA would be more stable and efficient than RIA because the inner reliability algorithms used to solve the first-order quantile approximation (the so-called \emph{Mean Value} algorithms) are argued to be much more efficient than their probability approximation counterparts (\eg the Hasofer-Lind-Rackwitz-Fiessler algorithm).\par

The nested approach becomes intractable in the case of more complex performance functions for which the nested reliability analysis should resort to simulation-based methods. For such cases however, \citet{Royset2004,Royset2004a} proposed the so-called \emph{sample average approximation} method which consists in gradually refining the simulation-based reliability analysis as the optimization algorithm converges towards an optimal design. To do so, they propose an empirical stepwise refinement criteria to define whether the number of simulations should be raised or not. Note also that the use of variance reduction techniques can significantly reduce the overall number of simulation runs such as demonstrated in \citet{Jensen2009} where the authors used subset simulation. However, simulation-based approaches still require too many performance functions evaluations to make the approach applicable to real engineering cases even though their application field is growing with the increasing availability of \emph{high performance computational} (HPC) resources (\eg interconnected clusters of PCs).\par

An alternative to double-loop approach consists in decoupling the optimization loop from the reliability analyses so that both can be sequentially performed in an independent manner. Such approaches are referred to as \emph{sequential approaches} or \emph{decoupled approaches} \citep{Royset2001,Du2004,Aoues2010}. An advantage of these approaches is that they do not require any reliability sensitivity analysis since the optimization is performed directly onto the performance function. However, the decoupling often relies on the \emph{most probable failure point} (MPFP) assumptions and thus suffers from the possible non-unicity of this point and the strong nonlinearities in the performance functions.\par

\emph{Single-loop approaches} \citep{Kuschel1997,KirjnerNeto1998,Kharmanda2002,Shan2008} attempt to fully reformulate the original RBDO problem into an equivalent DDO problem that allow\added{s} a simple and efficient resolution by means of classical optimization algorithms. The approach is mainly based on concepts that are closely related to the notion of partial safety factors. It is certainly the most computationally efficient approach as soon as the assumptions under which the probabilistic-deterministic equivalence is built hold. Once again, most of them are based upon the assumption that the MPFP exists and that it is unique.\par

\emph{Stochastic subset optimization} (SSO) is a simulation-based approach recently proposed by \citet{Taflanidis2009a} that consists in finding the region of the admissible design space where the failure probability density function is minimal. It is based on conditional simulations in a so-called \emph{augmented reliability space} where the design variables are artificially considered as random with uniform distribution. The range of uncertainty in the design variables is reduced along with the identification of low failure probability density regions. The overall concept is closely related with the subset simulation reliability method proposed by \citet{Au2001}, and further explored by \citet{Au2005} for reliability-based design sensitivity analysis. In the end, the algorithm provides a set of parameters that is likely to contain the optimal solution which can possibly be found by means of a more refined stochastic search algorithm. However, the problem that SSO attempts to solve is not a full RBDO problem in the sense that it is designed to minimize the failure probability whereas the purpose of RBDO is to minimize a cost function under some maximal failure probability constraint.\par

Finally, the approach that is investigated here is the \emph{surrogate-based} (or \emph{response-surface-based}) approach. Starting with the premise that the performance function evaluation might involve a time-consuming computational task, the approach consists in replacing this function by a surrogate that is much faster to evaluate. This surrogate is usually built and possibly refined from a few evaluations of the real performance function. Various surrogates have been used in the structural reliability literature out of which: polynomial response surfaces \citep{Faravelli1989,Bucher1990,Kim1997,Das2000}, polynomial chaos expansions \citep{Sudret2002,Berveiller2006a,BlatmanCras2008,BlatmanPEM2010}, support vector machines \citep{Hurtado2004,Deheeger2008,Bourinet2010}, neural networks \citep{Papadrakakis2002} and kriging \citep{Bichon2008}. \removed{The latter is} \added{Such surrogates are} argued here to be more flexible than the classical Taylor-expansion-based methods (\ie approaches using the first- or second-order reliability theory) that are commonly used in the RBDO literature. An additional interesting fact about this approach is that some of these surrogates, namely kriging and support vector machines, allow one to quantify an empirical surrogate error measure that can be propagated to the final quantity of interest: the failure probability.\par

\section{\label{sec:Kriging}The kriging surrogate}

In computer experiments, \emph{surrogate modeling} addresses the problem of replacing a time-consuming mathematical model $\cm$ called a \emph{simulator} by an \emph{emulator} that is much faster to evaluate. The emulator has the same input space $\cd_{\ve{x}} \subseteq \Rr^n$ and output space $\cd_{y} \subseteq \Rr$ than the simulator. For the sake of simplicity only scalar-output simulators are considered here. It is important to note here that both $\ve{x}$ and $y$ are deterministic in this context. In the present RBDO application, the time-consuming models are the performance functions $\{g_l,\,l=1\enu n_p\}$ in \eqref{eq:RBDO_RIA}.\par

Kriging \citep{Santner2003} is one particular emulator that is able to give a probabilistic response $\widehat{Y}\left(\ve{x}\right)$ whose variance (spread) depends on the quantity of available knowledge. In other words, the uncertainty in this prediction is purely epistemic and due to a lack of knowledge at specific input $\ve{x}$. For a more detailed discussion on the distinction that should be made between \emph{aleatory} and \emph{epistemic} sources of uncertainty, the reader is referred to \citet{DerKiureghian2009}. This interesting property of the kriging surrogate will be used in the next sections to make the surrogate-based RBDO approach adaptive.\par

In essence, kriging starts with the assumption that the output $y \equiv \cm\left(\ve{x}\right)$ is a sample path of a Gaussian stochastic process $Y$ whose mean and autocovariance functions are \emph{unknown} and to be determined from the values of the model response $\cy = \{y_i=\cm\left(\ve{x}_i\right),\;i=1\enu m\}$ evaluated onto an experimental design $\cx = \{\ve{x}_1\enu\ve{x}_m\}$. More specifically, the so-called \emph{universal kriging} (UK) model assumes the Gaussian process has a stationary autocovariance function and is centered with respect to a non-stationary trend in the form of a linear regression model so that it reads:%
\begin{equation} \label{eq:staGP}
  Y\left(\ve{x}\right) = \sum_{i=1}^p \beta_i\,f_i\left(\ve{x}\right) + Z\left(\ve{x}\right) = \ve{f}\left(\ve{x}\right)\tr\,\ve{\beta} + Z\left(\ve{x}\right).
\end{equation}%
The first term is the regression part that requires the choice of a functional set $\ve{f} \in \ldeu{\cd_{\ve{x}}}{\Rr}$. The second term is a stationary Gaussian process with zero mean, constant variance $\sigma_Y^2$ and stationary autocorrelation function $R$ so that its autocovariance function reads as follows:%
\begin{equation}
  C\left(\ve{x},\,\ve{x}'\right) = \sigma_Y^2\,R\left(\left|\ve{x}-\ve{x}'\right|,\,\ve{\ell}\right).
\end{equation}%
The parameters $\ve{\ell}$, $\ve{\beta}$ and $\sigma_Y^2$ shall be determined according to a given class of symmetric positive definite autocorrelation functions and a given set of regression functions. The most widely used class of autocorrelation functions is the anisotropic generalized exponential model:%
\begin{equation}
  R\left(\left|\ve{x}-\ve{x}'\right|,\ve{\ell}\right) = \exp\left(\sum\limits_{k=1}^n - \frac{\left|x_k-x_k'\right|}{\ell_k}^s\right), \quad 1 \leq s \leq 2.
\end{equation}%
Generally speaking, the choice of the autocorrelation model should be made consistently with the known properties of $\cm$ such as derivability or periodicity \citep[see \eg][ch. 4]{Rasmussen2006}. Common regression models involve zero, first and second-order polynomials of $\{x_k,\,k=1,\,\ldots,\,n\}$. Note however that as for ordinary least-squares regression, the size $p$ of the functional basis together with the dimension of the input vector $n$ dramatically increases the number $m$ of required observations. When attempting to model the output of a high-fidelity computer model, a good choice for the regression model may be an equivalent low-fidelity analytical model built on simplifying assumptions.\par

The construction of a kriging model consists in a two-stage framework that is further detailed in \removed{this} Subsections \ref{sec:stage1} and \ref{sec:stage2}.\par

\subsection{\label{sec:stage1}The conditional distribution of the prediction}

The first stage aims at looking for the distribution of some prediction $\widehat{Y}(\ve{x}) = Y(\ve{x}) \left|\ve{Y}\right.$ of the Gaussian process (GP) at any location $\ve{x} \in \cd_{\ve{x}}$ conditional on the vector of observations $\ve{Y} = \langle Y\left(\ve{x}_1\right)\enu Y\left(\ve{x}_m\right)\rangle\tr$. Since $Y$ is assumed to be a Gaussian process, the vector of observations $\ve{Y}$ is distributed as follows:
\begin{equation} \label{eq:Y}
  \ve{Y} \sim \cn\left(\ma{F}\,\ve{\beta},\;\sigma_Y^2\,\ma{R}\right)
\end{equation}%
where we have introduced the matrices $\ma{F}$ and $\ma{R}$ that are defined with respect to the assumed statistics of the Gaussian process. Their terms read as follows:
\begin{eqnarray}
  F_{ij} & = & f_j\left(\ve{x}_i\right), \quad i=1\enu m,\;j=1\enu p \\
  R_{ij} & = & R\left(\left|\ve{x}_i - \ve{x}_j\right|,\,\ve{\ell}\right), \quad i=1\enu m,\;j=1\enu m.
\end{eqnarray}
It is also known that the augmented random vector $\left\langle\ve{Y}\tr,\,Y\left(\ve{x}\right)\right\rangle\tr$ is jointly Gaussian by definition:
\begin{equation} \label{eq:YYx}
  \left\{\begin{array}{c}
    \ve{Y} \\
    Y\left(\ve{x}\right)
  \end{array}\right\}
  \sim \cn\left(
  \left\{\begin{array}{c}
    \ma{F} \\
    \ve{f}\left(\ve{x}\right)\tr
  \end{array}\right\}\,\ve{\beta},\;
  \sigma_Y^2\,\left[\begin{array}{cc}
    \ma{R} & \ve{r}\left(\ve{x}\right) \\
    \ve{r}\left(\ve{x}\right)\tr & 1
  \end{array}\right]
  \right)
\end{equation}%
where we have introduced the vector of cross-correlations $\ve{r}(\ve{x})$ between the observations and the prediction:
\begin{equation}
  r_i(\ve{x}) = R\left(\left|\ve{x} - \ve{x}_i\right|,\,\ve{\ell}\right), \quad i=1\enu m.
\end{equation}\par

It is well known \citep[see \eg][Chap. 8]{Severini2005} that the conditional distribution of $\widehat{Y}(\ve{x}) = Y(\ve{x})\mid\ve{Y}$ is also Gaussian with mean:
\begin{equation} \label{eq:pred_mean}
  \mu_{\widehat{Y}}\left(\ve{x}\right) = \ve{f}\left(\ve{x}\right)\tr\,\widehat{\ve{\beta}}
                                       + \ve{r}\left(\ve{x}\right)\tr\ma{R}^{-1}\left(\ve{Y} - \ma{F}\,\widehat{\ve{\beta}}\right)
\end{equation}%
and variance:%
\begin{equation} \label{eq:pred_var}
    \sigma_{\widehat{Y}}^2\left(\ve{x}\right) = \sigma_{Y}^2\,\left(
    1
    - \ve{r}(\ve{x})\tr\,\ma{R}^{-1}\,\ve{r}(\ve{x})
    + \ve{u}(\ve{x})\tr\,(\ma{F}\tr\,\ma{R}^{-1}\,\ma{F})^{-1}\,\ve{u}(\ve{x})
    \right)
\end{equation}%
where we have introduced:
\begin{eqnarray}
    \widehat{\ve{\beta}} & = & (\ma{F}\tr\,\ma{R}^{-1}\,\ma{F})^{-1}\,\ma{F}\tr\,\ma{R}^{-1}\,\ve{Y} \\
    \ve{u}(\ve{x}) & = & \ma{F}\tr\,\ma{R}^{-1}\,\ve{r}(\ve{x}) - \ve{f}(\ve{x}).
\end{eqnarray}%

\subsection{Properties of the kriging surrogate}

The kriging surrogate is an exact interpolator. Indeed, noting that the $i$-th column of the matrix $\ma{R}$ is equal to the vector $\ve{r}\left(\ve{x}_i\right)=\ve{e}_i$, the following relation holds:%
\begin{equation}
\ma{R}\,\ve{e}_i = \ve{r}\left(\ve{x}_i\right)
\end{equation}%
where $e_i$ is the $i$-th basis vector of $\Rr^m$ which has all-zero components except its $i$-th component equal to one. Using the symmetry property of $\ma{R}$ and plugging the latter relation in the predictor's mean in \eqref{eq:pred_mean} and in its variance in \eqref{eq:pred_var} finally allows to prove the following statement:%
\begin{equation}
\left\{\begin{array}{rcl}
  \mu_{\widehat{Y}}\left(\ve{x}_i\right)    & = & y_i \\
  \sigma_{\widehat{Y}}\left(\ve{x}_i\right) & = & 0
\end{array}\right. \quad i = 1\enu m, \quad \forall \ve{\ell}, \ve{\beta}, \sigma_Y^2
\end{equation}%
which means that the kriging surrogate interpolates the observations with variance equal to zero (deterministic prediction). This property holds for any regression and regular autocorrelation models, whatever the values of the parameters $\ve{\ell}$, $\beta$ and $\sigma_Y^2$.\par

\subsection{\label{sec:stage2} Joint maximum likelihood estimation of the Gaussian process parameters}

In the first stage it was assumed that the autocovariance function $\sigma_Y^2\,R\left(\bullet,\,\ve{\ell}\right)$ and the regression model $\ve{f}\left(\bullet\right)\tr\,\ve{\beta}$ are \emph{known}. In \emph{computer experiments} or in \emph{geostatistics} it is never the case so that the models need to be chosen and their parameters must be determined using common statistical inference techniques such as the \emph{variographic analysis} (VA), \emph{maximum likelihood estimation} (MLE) or \emph{Bayesian estimation} (BE).

Historically, the kriging methodology was introduced by geostatisticians in order to predict a map of soil properties from \emph{in situ} observations for mining prospection. In this field, the autocovariance structure is usually estimated from the data using \emph{variographic analysis} \citep{Cressie1993}. Then, provided the \emph{empirical variogram} features some required properties it can be turned into an autocovariance model. However, this methodology is not well-suited to our purpose because of its user-interactivity.

In computer experiments, the most widely used methodology is the MLE technique. Provided a functional set $\ve{f} \in \ldeu{\cd_{\ve{x}}}{\Rr}$ and a stationary \removed{covariance} \added{autocorrelation} model $R(\bullet,\,\ve{\ell})$ are chosen, one can express the likelihood of the data with respect to the model and maximize it with respect to the sought parameters ($\ve{\ell}$, $\ve{\beta}$ and $\sigma_Y^2$). One can show that $\ve{\beta}$ and $\sigma_Y^2$ can be derived analytically (using the first-order optimality conditions) and solely depends on the autocovariance parameters $\ve{\ell}$ that are solution of a numerically tractable global optimization problem -- see \eg \citet{Welch1992, Lophaven2002} for more details. This technique, implemented within the DACE toolbox by \citet{Lophaven2002}, was used for the applications presented in this paper.

Note that the BE technique consists in weighting the likelihood by a prior over the autocovariance parameters and finally allows to derive a full probabilistic posterior on them. Despites it has an attractive formulation as it allows one to quantify an additional epistemic uncertainty in the autocovariance parameters themselves, it is still difficult to involve it in a nested general-purpose procedure such as the one proposed in this paper due to the required prior information which might not be available.

In the case of MLE, the variance $\sigma_{\widehat{Y}}^2$ underestimates the real variance as it does not account for \textit{(i)} the empirical choice of the autocovariance structure of the GP, and \textit{(ii)} the uncertainty induced by the statistical inference of $\ve{\widehat{\ell}}$. Note that BE does account for this uncertainty but it is rather difficult to propagate this additional uncertainty through the predictor -- this is sometimes referred to as \emph{Bayesian kriging}. Even though, the probabilistic response of the kriging model is convenient as it allows one to:
\begin{itemize}
 \item provide probabilistic confidence intervals in addition to the prediction;
 \item define refinement criteria to build the kriging surrogate in an adaptive refinement procedure depending on the region of interest as introduced in Section~\ref{sec:DOE}.
\end{itemize}

\section{\label{sec:DOE}Design of experiments for the kriging surrogate}

Various refinement techniques have been proposed in the kriging-related literature, \eg for global optimization \citep{Jones1998} or for probability/quantile estimation \citep{Oakley2004b,Bichon2008,Lee2008,Ranjan2008,Vazquez2009,Picheny2010b}. They are all based on the genuine probabilistic nature of the kriging prediction. An adaptive refinement technique consists in finding the set of points that should be added to the \emph{design of experiments} (DOE) in order to improve the prediction and reduce its associated epistemic uncertainty in some specific \emph{region of interest} \citep{Picheny2010b}. In global optimization, the region of interest is the vicinity of the optimum that can be sequentially achieved by means of the so-called \emph{efficient global optimization} technique \citep{Jones1998}. In reliability and reliability-based design optimization, the region of interest in the space of random variables $\ve{X}$ is the vicinity of the limit-state surface $g(\ve{x})=0$ (we drop the subscript $l$ in this section for the sake of clarity).\par

\subsection{Premise}

Provided an initial kriging prediction of the performance function $g$ denoted by $\widehat{G} \sim \cn\left(\mu_{\widehat{G}},\,\sigma_{\widehat{G}}\right)$, the probable vicinity of the limit-state surface $g=0$ can be defined in terms of the following \emph{margin of uncertainty} (epistemic uncertainty):%
\begin{equation}
 \mathbb{M} = \acc{\ve{x}\,:\,-k\,\sigma_{\widehat{G}}\left(\ve{x}\right) \leq \mu_{\widehat{G}}\left(\ve{x}\right) \leq +k\,\sigma_{\widehat{G}}\left(\ve{x}\right)}
\end{equation}%
where $k$ might be chosen as $k=\Phi^{-1}\left(97.5\%\right)=1.96$ meaning that a 95\% confidence interval onto the prediction of the limit-state surface is chosen. Any point that belongs to this margin correspond to an uncertain sign of the prediction. The spread of this margin should be reduced in order to make the reliability estimation accurate, and this can be achieved by adding new points in the DOE that are located in this margin.\par

The probability that a point $\ve{x}\in\cd_{\ve{x}}$ falls into the margin $\mathbb{M}$ (with respect to the epistemic uncertainty in the prediction) can be expressed in closed-form as follows:%
\begin{equation}\label{eq:P_x_in_M}
 \Prob{\ve{x}\in\mathbb{M}} = \Phi\left(\frac{k\,\sigma_{\widehat{G}}\left(\ve{x}\right) - \mu_{\widehat{G}}\left(\ve{x}\right)}{\sigma_{\widehat{G}}\left(\ve{x}\right)}\right) - \Phi\left(\frac{-k\,\sigma_{\widehat{G}}\left(\ve{x}\right) - \mu_{\widehat{G}}\left(\ve{x}\right)}{\sigma_{\widehat{G}}\left(\ve{x}\right)}\right).
\end{equation}%
Note that the same criterion is used in \citet{Picheny2010b} where it has been named the \emph{weighted integrated mean squared error} (W-IMSE). Then, finding the point that maximizes this quantity on the support of the PDF of $\ve{X}$ will finally bring the best improvement point in the DOE. Starting with this statement, many authors in the kriging literature decide to use global optimization algorithms in order to find \emph{the best} improvement point. For instance, \citet{Bichon2008} use a different criterion named the \emph{expected feasibility function}, and \citet{Lee2008} use the \emph{constraint boundary sampling} criterion, but both use a global optimization algorithm to find the best improvement point.\par

\subsection{Principle of the proposed refinement procedure}

In this paper, we propose a different strategy that allows one to add \emph{a set} of improvement points. Indeed, all the proposed refinement criteria (including the one in \eqref{eq:P_x_in_M}) are highly multimodal, thus meaning that:
\begin{itemize}
    \item the global optimization problem is not easy to solve;
    \item there does not really exist \emph{one} best point (especially for the contour approximation problem of interest), since each mode is associated with a potentially interesting point to add to the DOE.
\end{itemize}
It may also be faster if one has the ability to perform several simulation runs of $g$ at the same time (\eg on some distributed computational platform). Note that the idea is inspired from \citet{Hurtado2004,Deheeger2007,Deheeger2008,Bourinet2010} where the authors use the same idea applied to \emph{support vector classifiers}.\par

To do so, the probability to fall in the margin of epistemic uncertainty is multiplied by a weighting PDF $w$ so that the following criterion:%
\begin{equation} \label{eq003}
 \cc\left(\ve{x}\right) = \Prob{\ve{x}\in\mathbb{M}} \, w\left(\ve{x}\right)
\end{equation}%
can itself be considered as a PDF up to an unknown but finite normalizing constant. The weighting density $w$ can either be the original PDF of $\ve{X}$, or, as it is proposed here, the uniform PDF on a sufficiently large confidence region of the original PDF.\par

Such a confidence region might be difficult to define for any given PDF, but as it is usually done in structural reliability \citep{Ditlevsen1996}, the original random vector $\ve{X}$ can be transformed into a probabilistically equivalent standard Gaussian random vector $\ve{U}$ for which the confidence region is simply an hypersphere with radius $\beta_0$. The reader is referred to \citet{Lebrun2009a} for a recent discussion on such mappings $\ve{u} = T(\ve{x})$. In that given space, $\beta_0$ can be easily selected as \eg $\beta_0=8$ which corresponds to the maximal reliability index that can be justified numerically, and the sought uniform PDF is simply defined in terms of the following indicator function:%
\begin{equation}
  w\left(\ve{u}\right) \propto \fcar{\sqrt{\ve{u}\tr\,\ve{u}} \leq \beta_0}{\ve{u}}.
\end{equation}%
The normalizing constant of this \removed{improper} PDF is the volume of the hypersphere and it could thus be easily derived though it is not required to generate samples from $\cc$ as explained hereafter. \added{Note that another weighting PDF will be introduced later on (in Section \ref{sec:AugmentedReliabilitySpace}) for the RBDO problem of interest.}\par

Given that pseudo-PDF, one can generate $N$ (say $N=10^4$) samples by means of any well suited \emph{Markov chain Monte Carlo} (MCMC) simulation technique \citep{Robert2004}, such as the \emph{slice sampling} technique \citep{Neal2003}. The $N$ generated candidates are expected to be highly concentrated at the modes of $\cc$ -- and thus in $\mathbb{M}$. This population of candidates is then reduced to a smaller one that has essentially the same statistical properties (\ie it uniformally covers $\mathbb{M}$) by means of the $K$-means clustering technique \citep{MacQueen1967}.\par

We make a final remark considering the three approximate failure subsets $\widehat{F}^{-1}$, $\widehat{F}^{~0}$, $\widehat{F}^{+1}$ defined as follows:%
\removed{
\begin{equation}
 \widehat{F}^i = \acc{\ve{x}\;:\;\widehat{G} + i\,k\,\sigma_{\widehat{G}}\left(\ve{x}\right) \leq 0}, \quad i = -1,\,0,\,+1.
\end{equation}%
}
\added{
\begin{equation}
 \widehat{F}^i = \acc{\ve{x}\;:\;\mu_{\widehat{G}}(\ve{x}) + i\,k\,\sigma_{\widehat{G}}\left(\ve{x}\right) \leq 0}, \quad i = -1,\,0,\,+1.
\end{equation}%
}
Note that the extreme failure subsets are related with the margin of uncertainty through the following set relationship: $\mathbb{M} = \widehat{F}^{-1} \setminus \widehat{F}^{+1}$. Due to the positiveness of standard deviation, the following statement holds:%
\begin{equation}
 \widehat{F}^{+1} \subseteq \widehat{F}^{~0} \subseteq \widehat{F}^{-1} \Rightarrow \Prob{\ve{X}\in\widehat{F}^{+1}} \leq \Prob{\ve{X}\in\widehat{F}^{~0}} \leq \Prob{\ve{X}\in\widehat{F}^{-1}}.
\end{equation}%
Unfortunately, there is no proof that the real failure subset $F$ satisfies the following relationship $\widehat{F}^{+1} \subseteq F \subseteq \widehat{F}^{-1}$ due to the empirical assumptions that were made in Section~\ref{sec:Kriging}. However, the spread of the interval $[\Pro(\ve{X}\in\widehat{F}^{+1}),\;
\Pro(\ve{X}\in\widehat{F}^{-1})]$ is a useful measure to check if the kriging surrogate is accurate enough for reliability analysis or not, and it is used in this paper as a stopping criterion for the proposed refinement procedure. Note that due to the inconvenient order of magnitude of low probabilities, it is more meaningful to work with the generalized reliability indices that are defined as follows:
\begin{equation}
  \widehat{\beta}^{~i} = -\Phi^{-1}\left(\Pro(\ve{X}\in\widehat{F}^{~i})\right), \quad i=-1,\,0,\,+1.
\end{equation}%
Then, in the proposed applications, the accuracy criterion is usually set to $\varepsilon_{\beta} = 10^{-1}-10^{-2}$ and the refinement stops when:
\begin{equation}
  \max\left(\widehat{\beta}^{+1} - \widehat{\beta}^{~0},\,\widehat{\beta}^{~0} - \widehat{\beta}^{-1}\right) \leq \varepsilon_{\beta}.
\end{equation}%
When $\widehat{\beta}^{~i}$ are estimated by means of simulation techniques (Monte-Carlo \removed{simulation} or subset simulation), the proposed accuracy criterion should account for the additional uncertainty induced by the lack of simulations. To do so, in the present paper, $\widehat{\beta}^{-1}$ and $\widehat{\beta}^{+1}$ estimates are replaced with their respective lower and upper 95\% confidence bounds based on their associated \emph{variance of estimation}. Thus, $\varepsilon_{\beta}$ should be selected in accordance with the given number of simulations used for reliability estimation.

\subsection{Implementation}

In order to summarize the proposed refinement procedure, we provide the pseudo-code in Algorithm~\ref{alg:DOE}. First, we initialize the empty DOE $\left(\cx = \emptyset, \cg = \emptyset\right)$, the uniform refinement pseudo-PDF $\cc$ for the first space-filling DOE and we select the level of confidence $k$ in the metamodel. Then, we generate the candidate population $\cp$ from the \removed{improper} density function $\cc$ by means of any well-suited MCMC simulation technique (using \eg slice sampling). This population is reduced to its $K$ clusters' center \added{using $K$-means clustering} -- $K$ being given. The performance function $\cg$ is evaluated onto these $K$ newly selected points $\cx_{\rm new}$ and a new kriging model is built from the updated DOE $\left(\cx, \cg\right)$. \added{Note that the kriging model construction step involves the maximum likelihood estimation of the autocorrelation parameters $\ve{\ell} \in [\ve{\ell}_L, \ve{\ell}_U]$.} The refinement pseudo-PDF $\cc$ is also updated. Finally, a reliability analysis (using \eg subset simulation) is performed onto the three approximate failure subsets $\widehat{F}^{+1}$, $\widehat{F}^{~0},$ and $\widehat{F}^{-1}$ in order to compute the proposed error measure. The DOE is enriched if and while this error measure exceeds a given tolerance $\varepsilon_{\beta} = 10^{-1}-10^{-2}$.\par

\begin{algorithm}[H]
  \caption{Population-based adaptive refinement strategy}
  \label{alg:DOE}
  \begin{algorithmic}[1]
%     \STATE \COMMENT{Initializations}
    \STATE $\cx = \emptyset$, $\cg = \emptyset$
    \STATE $w := \ve{x} \mapsto w(\ve{x})$
    \STATE $k := \Phi^{-1}(97.5\%)$, $\varepsilon_{\beta} := 10^{-1}$
%     \STATE \COMMENT{Refinement loop}
    \STATE Refine := \TRUE
    \STATE $\Prob{\ve{x}\in\mathbb{M}} := \ve{x} \mapsto 1$
    \WHILE {Refine}
%       \STATE \COMMENT{1. Sample a candidate population from this PDF}
      \STATE $\cp := {\rm MCMCAlgorithm}(\cc)$
%       \STATE \COMMENT{2. Reduce the candidate population to its $K$ clusters' center}
      \STATE $\cx_{\rm new} := {\rm KMeansAlgorithm}(\cp,\,K)$
%       \STATE \COMMENT{3. Evaluate the performance function and enrich current DOE}
      \STATE $\cg_{\rm new} := g\left(\cx_{\rm new}\right)$
      \STATE $\cx := \{\cx,\,\cx_{\rm new}\}$, $\cg := \{\cg,\,\cg_{\rm new}\}$
%       \STATE \COMMENT{4. Build the new kriging model}
      \STATE $\widehat{G} := {\rm MaximumLikelihoodKrigingModel}(\cx,\,\cg,\,\ve{f}\left(\bullet\right),\,R\left(\bullet,\,\ve{\ell}\right),\,\ve{\ell}_L,\,\ve{\ell}_U)$
%       \STATE \COMMENT{5. Update the \emph{margin of uncertainty}}
      \STATE $\Prob{\ve{x}\in\mathbb{M}} := \ve{x} \mapsto \Phi\left(\frac{k\,\sigma_{\widehat{G}}\left(\ve{x}\right) - \mu_{\widehat{G}}\left(\ve{x}\right)}{\sigma_{\widehat{G}}\left(\ve{x}\right)}\right) - \Phi\left(\frac{-k\,\sigma_{\widehat{G}}\left(\ve{x}\right) - \mu_{\widehat{G}}\left(\ve{x}\right)}{\sigma_{\widehat{G}}\left(\ve{x}\right)}\right)$
%       \STATE \COMMENT{6. Define the refinement pseudo-PDF}
      \STATE $\cc := \ve{x} \mapsto \Prob{\ve{x}\in\mathbb{M}} \, w(\ve{x})$
%       \STATE \COMMENT{7. Perform reliability analyses on the three failure subsets}
      \FOR {$i := -1,\,0,\,+1$}
	\STATE $\widehat{F}^{~i} := \acc{\ve{x}~:~\mu_{\widehat{G}}(\ve{x}) + i\,k\,\sigma_{\widehat{G}}(\ve{x}) \leq 0}$
	\STATE $\widehat{\beta}^{~i} := {\rm ReliabilityAnalysis}(\widehat{F}^{~i})$
      \ENDFOR
%       \STATE \COMMENT{8. Refine or done?}
      \STATE ${\rm Refine} := \max\left(\widehat{\beta}^{+1} - \widehat{\beta}^{~0},\,\widehat{\beta}^{~0} - \widehat{\beta}^{-1}\right) > \varepsilon_{\beta}$
    \ENDWHILE
  \end{algorithmic}
\end{algorithm}

\begin{figure}[t]
  \centering
    \subfigure[Init.: 10-point DOE]{\label{fig:Waarts_unifbeta_init}
    \begin{minipage}{.3\textwidth}\begin{center}
      \includegraphics[width=\textwidth, clip=true, trim = 0 0 255 0]{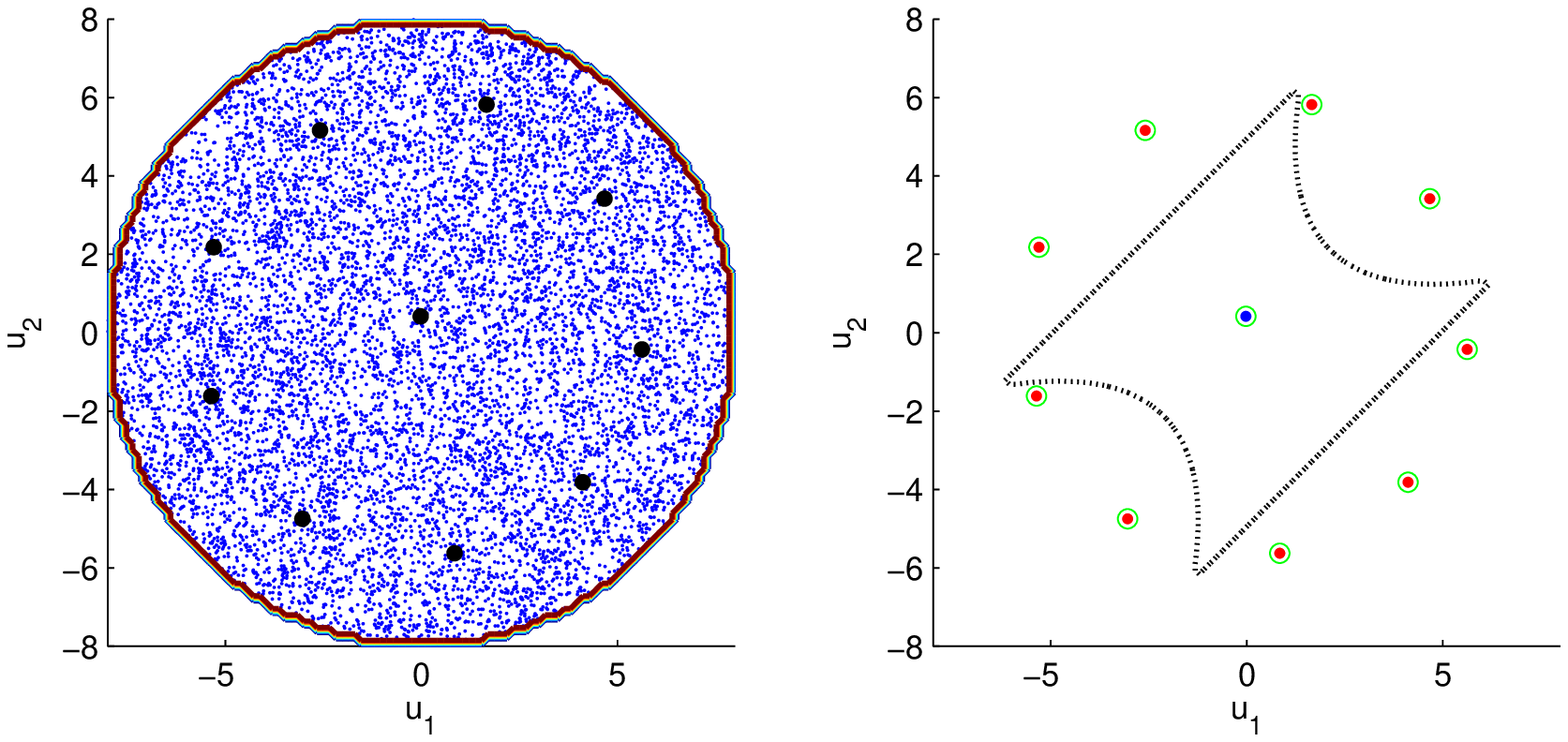} \\
      \includegraphics[width=\textwidth, clip=true, trim = 255 0 10 0]{Waarts_unifbeta_1}
    \end{center}\end{minipage}}
    \subfigure[Step 3: 40-point DOE]{\label{fig:Waarts_unifbeta_step3}
    \begin{minipage}{.3\textwidth}\begin{center}
      \includegraphics[width=\textwidth, clip=true, trim = 0 0 255 0]{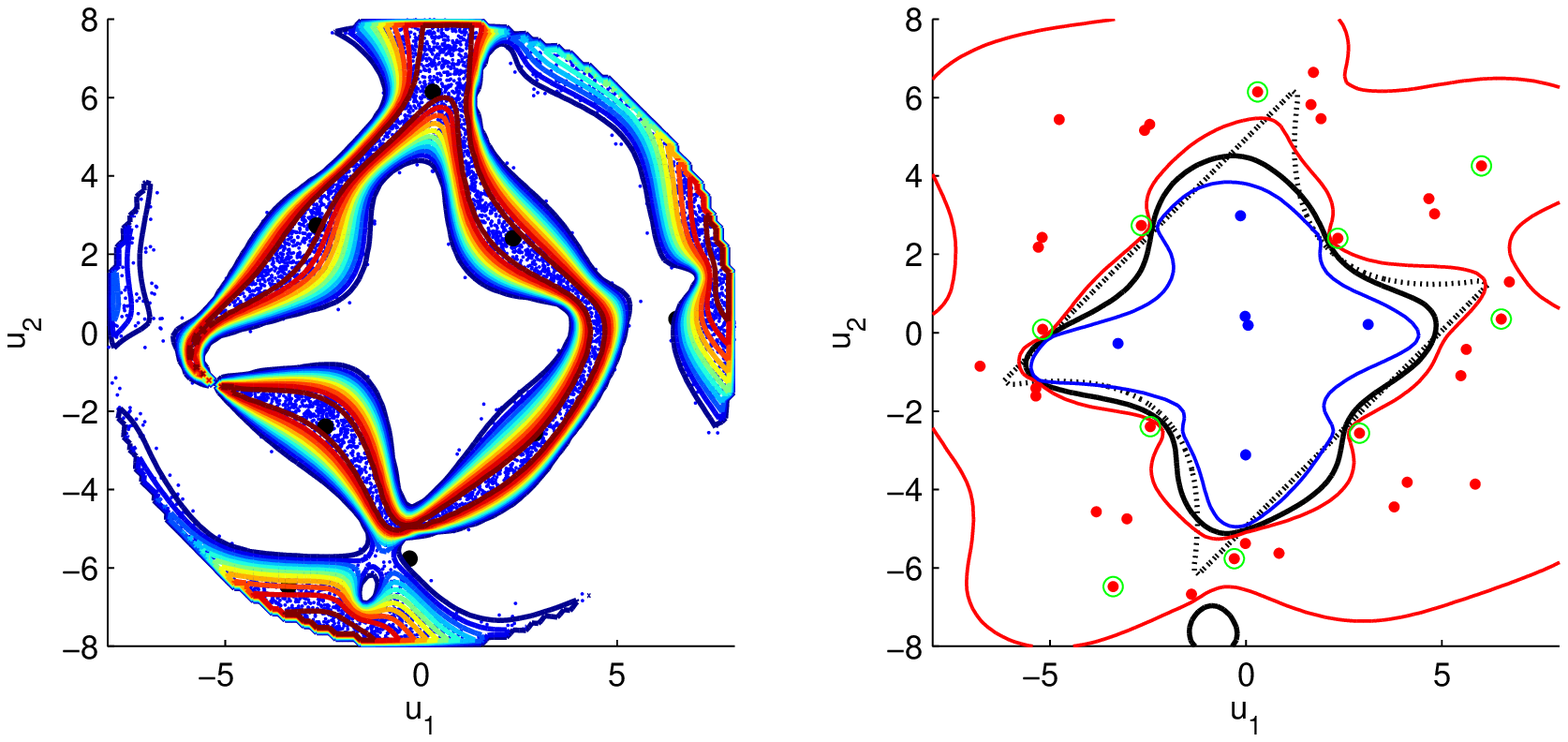} \\
      \includegraphics[width=\textwidth, clip=true, trim = 255 0 10 0]{Waarts_unifbeta_4}
    \end{center}\end{minipage}}
    \subfigure[Step 9: 100-point DOE]{\label{fig:Waarts_unifbeta_step9}
    \begin{minipage}{.3\textwidth}\begin{center}
      \includegraphics[width=\textwidth, clip=true, trim = 0 0 255 0]{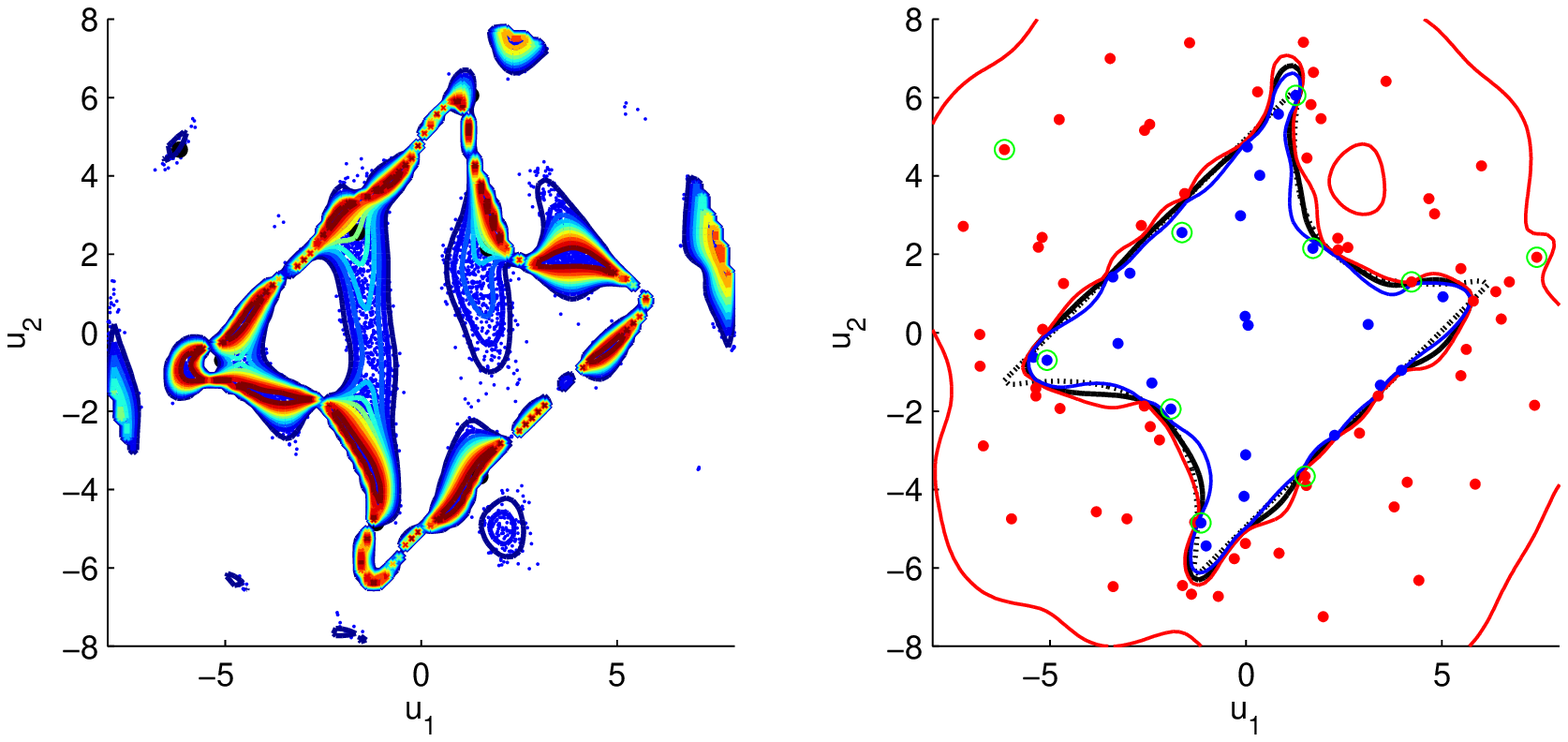} \\
      \includegraphics[width=\textwidth, clip=true, trim = 255 0 10 0]{Waarts_unifbeta_9}
    \end{center}\end{minipage}}
  \caption{An illustration of the proposed limit-state surface refinement technique}
  \label{fig:DOE_refinement}
\end{figure}

Figure~\ref{fig:DOE_refinement} illustrates the proposed adaptive refinement strategy applied to a nonlinear limit-state surface from \citet{Waarts2000}. The upper subfigures show the contours of the refinement pseudo-PDF $\cc$ at each refinement step together with the candidate population generated by slice sampling and its $K$ clusters' center \added{obtained by $K$-means clustering} -- $K=10$ being given. \added{For this application, the weighting PDF $w$ was selected as the uniform density in the $\beta_0$-radius hypersphere -- $\beta_0=8$. It can be observed that the refinement criterion features several modes as argued earlier in this Section.} In the lower subfigures one can see the real limit-state surface represented as the dashed black curve, its kriging prediction represented as the black line and its associated margin of uncertainty $\mathbb{M}$ which is bounded below by the red line and above by the blue line. Another interpretation of these figures is that any point within the blue bounded shape is positive with a 95\% confidence and any point inside the red bounded shape is negative with the same confidence level.\par

\section{\label{sec:HSBRBDO}The proposed adaptive surrogate-based RBDO strategy}

The proposed strategy to solve the RBDO problem in \eqref{eq:RBDO_RIA} consists in nesting the previously introduced kriging surrogate together with the proposed refinement strategy within a classical (but efficient) nested RBDO algorithm.\par

\subsection{\label{sec:AugmentedReliabilitySpace}The augmented reliability space}

In this section, we describe the space where the kriging surrogates are built. Indeed, observing that building the kriging surrogates from \removed{from} an empty DOE for each nested reliability analysis (\eg in the space of the standard normal random variables) would be particularly inefficient, it is proposed to build and refine one unique \emph{global} kriging surrogate for all the nested reliability analyses.

\begin{figure}
  \centering
  \includegraphics[width=.7\textwidth, clip=true, trim = 65 35 65 35]{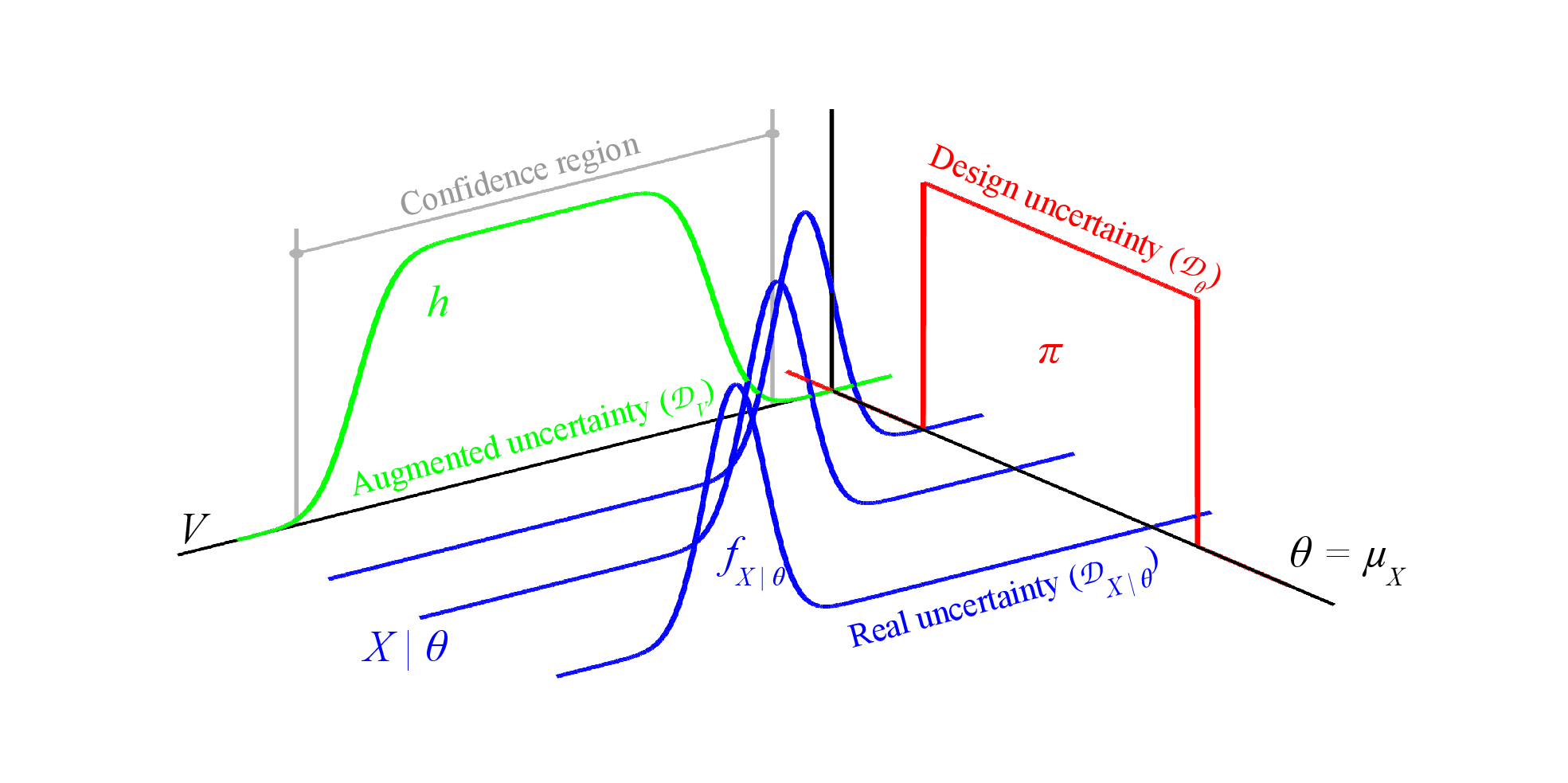}
  \caption{The augmented probability density function of a Gaussian random variate with uniform mean.}
  \label{fig:AugmentedPDF}
\end{figure}

Such a globality can be achieved by working in the so-called \emph{augmented reliability space} such as defined in \citet{Taflanidis2009}. In \citet{Kharmanda2002}, the augmented reliability space is defined as the tensor product between the space of the standardized normal random variables and the design space: $\cd_{\ve{U}} \times \cd_{\ve{\theta}}$, but the dimension of this space ($n+n_{\ve{\theta}}$) suffers from both the number of random variables $n$ and the number of design variables $n_{\ve{\theta}}$. It is also argued here that this space may cause some loss of information as the performance functions are not in bijection with that augmented space. In contrast, in \citet{Taflanidis2009} and in the present approach, the dimension of the augmented reliability space is kept equal to $n$ by considering that the design vector $\ve{\theta}$ simply augments the uncertainty in the random vector $\ve{X}$. Indeed, the augmented random vector $\ve{V} \equiv \ve{X}\left(\ve{\Theta}\right)$ has a PDF $h$ which accounts for both an \emph{instrumental uncertainty} in the design choices $\ve{\Theta}$ and the aleatory uncertainty in the random vector $\ve{X}$. Under such considerations, $h$ reads as follows:%
\begin{equation}
 h\left(\ve{v}\right) = \int_{\cd_{\ve{\theta}}} f_{\ve{X}}\left(\ve{x} \left| \ve{\theta} \right.\right)\,\pi\left(\ve{\theta}\right)\,\di{\ve{\theta}}
\end{equation}%
where $f_{\ve{X}}$ is the PDF of $\ve{X}$ given the parameters $\ve{\theta}$ and $\pi$ is the PDF of $\ve{\Theta}$ that can be assumed uniform on the design space $\cd_{\ve{\theta}}$. An illustration of this augmented PDF is provided in Figure~\ref{fig:AugmentedPDF} in the univariate case. The augmented reliability space is spanned by the axis $\ve{V}$ on the left in this simple case.

The DOE should cover uniformally a sufficiently large \emph{confidence region} of this augmented PDF in order to make the surrogate limit-state surfaces accurate wherever they can potentially be evaluated along the optimization process. More precisely, they should be accurate for extreme design choices $\ve{\theta}$ (\ie located onto the boundaries of the optimization space $\cd_{\ve{\theta}}$) and extreme values of the marginal random vector $\ve{X}$ (to be able to compute reliability indices as large as \eg $\beta_0=8$). A confidence region is essentially the multivariate extension of the univariate concept of \emph{confidence interval}. Under the previous general assumptions, it is hard to give a mathematical form to the contour of this region. However, one may easily build an hyperrectangular region that bounds the confidence region of interest.

Indeed, such an hyperrectangular region is defined as the tensor product of the confidence intervals on the augmented margins $\{V_i,\;i=1\enu n\}$. In order to compute the quantiles bounding these confidence intervals, one should additionally assume that the design parameters are exclusively involved in the definition of the margins -- \ie no parameters in the dependence structure (the \emph{copula}) as it will never be the case in most RBDO applications. For each margin, the lower quantiles $\{q_{V_i}^-,\,i=1\enu n\}$ (at the probability level $1-\Phi\left(\beta_0\right)=\Phi\left(-\beta_0\right)$) and upper quantiles $\{q_{V_i}^+,\,i=1\enu n\}$ (at the probability level $\Phi\left(+\beta_0\right)$) are respectively solutions of the following optimization problems:%
\begin{eqnarray}
 q_{V_i}^- & = & \min\limits_{\ve{\theta} \in \cd_{\ve{\theta}}} F_{X_i}^{-1}\left(\Phi(-\beta_0)\,\left|\,\ve{\theta}\right.\right), \quad i=1\enu n \\
 q_{V_i}^+ & = & \max\limits_{\ve{\theta} \in \cd_{\ve{\theta}}} F_{X_i}^{-1}\left(\Phi(+\beta_0)\,\left|\,\ve{\theta}\right.\right), \quad i=1\enu n
\end{eqnarray}
where $\{F_{X_i}^{-1},\;i=1\enu n\}$ are the quantile functions of the margins. If the domain $\cd_{\ve{\theta}}$ is rectangular and if one is able to derive an analytical expression for the quantile functions of the margins and their derivatives with respect to the parameters, then these optimization problems might be solved analytically. However assuming a more general setup where one has only numerical definition of these quantities, these problems can be efficiently solved by means of a simple gradient-based algorithm due to the convenient properties of the quantile functions -- namely, the monotony with respect to the location and shape parameters. Finally, the sought hyperrectangle can be easily defined by means of the following indicator function:%
\begin{equation} \label{eq:AugmentedWeightingPDF}
 \fcar{\cd_{\ve{V}},\,\beta_0}{\ve{v}} =
 \left\{\begin{array}{l}
  1\quad\text{if~} q_{V_i}^- \leq v_i \leq q_{V_i}^+,\quad i=1\enu n \\
  0\quad\text{otherwise}
 \end{array}\right.
\end{equation}%
Again, the normalizing constant of this \removed{improper} PDF could be easily derived (hyperrectangle volume) though it is not required by the refinement procedure proposed in Section~\ref{sec:DOE}.

\subsection{The adaptive surrogate-based nested RBDO algorithm}

\subsubsection{Optimization}

The kriging surrogate together with its adaptive refinement procedure is finally plugged into a double-loop RBDO algorithm. The outer optimization loop is performed by means of the Polak-He optimization algorithm \citep{Polak1997}. Provided an initial design, this algorithm proceeds iteratively in two steps: \textit{(i)} the direction of optimization is determined solving a quasi-SQP sub-optimization problem and \textit{(ii)} the step size is approximated by the Goldstein-Armijo approximate line-search rule.\par

\subsubsection{Surrogate-based reliability and reliability sensitivity analyses}

The nested reliability and reliability sensitivity analyses are performed with the \emph{subset simulation} variance reduction technique \citep{Au2001} onto the kriging surrogates. The subset simulation technique for reliability sensitivity analysis is detailed in \citet{Song2009}. Briefly, it takes advantage of the definition of the failure probability given in \eqref{eq:pfdef}. Indeed, pointing out that the limit-state equation does not explicit\removed{e}ly depend on the design variables $\ve{\theta}$, the differentiation of the failure probability only requires the differentiation of the joint PDF which can be derived analytically when the probabilistic model is defined in terms of margin distributions and copulas \citep{Lebrun2009a}. The trick is inspired from importance sampling and proceeds as follows:
\begin{eqnarray}
 \frac{\partial P_f}{\partial \theta} & \equiv & \frac{\partial}{\partial \theta} \int_{g\left(\ve{x}\right) \leq 0} f_{\ve{X}}\left(\ve{x},\,\ve{\theta}\right)\,\di{\ve{x}} \\
                                      &    =   & \int_{g\left(\ve{x}\right) \leq 0} \frac{\partial f_{\ve{X}}\left(\ve{x},\,\ve{\theta}\right)}{\partial \theta}\,\di{\ve{x}} \\
                                      &    =   & \int_{g\left(\ve{x}\right) \leq 0} \frac{\frac{\partial f_{\ve{X}}\left(\ve{x},\,\ve{\theta}\right)}{\partial \theta}}{f_{\ve{X}}\left(\ve{x},\,\ve{\theta}\right)}\,f_{\ve{X}}\left(\ve{x},\,\ve{\theta}\right)\,\di{\ve{x}} \\
				      & \equiv & \Espe{f_{\ve{X}}}{\fcar{g\leq0}{\ve{X}}\,\frac{\frac{\partial f_{\ve{X}}\left(\ve{X},\,\ve{\theta}\right)}{\partial \theta}}{f_{\ve{X}}\left(\ve{X},\,\ve{\theta}\right)}}.
\end{eqnarray}
The latter quantity is known to have an unbiased consistent estimator which reads as follows:
\begin{equation}
 \widehat{\frac{\partial P_f}{\partial \theta}} = \frac{1}{N} \sum\limits_{k=1}^N \fcar{g\leq0}{\ve{x}^{(k)}}\,\frac{\frac{\partial f_{\ve{X}}\left(\ve{x}^{(k)},\,\ve{\theta}\right)}{\partial \theta}}{f_{\ve{X}}\left(\ve{x}^{(k)},\,\ve{\theta}\right)}
\end{equation}
where the sample $\{x^{(1)}\enu x^{(N)}\}$ is the same as the one used for the estimation of the failure probability $P_f$. In other words, the estimation of $\widehat{\frac{\partial P_f}{\partial \theta}}$ does not require any additional simulation runs: it simply consists in a post-processing of the samples generated for reliability estimation. The concept can be easily extended to the subset simulation technique -- see \citet{Song2009} for the details.\par

\subsubsection{Implementation}

The overall methodology was implemented within the FERUM v4.0 toolbox \citep{Bourinet2009}. It makes use of the Matlab toolboxes functions \texttt{quadprog} (for the SQP sub-optimization problem) and \texttt{slicesample} (to generate samples from the refinement \removed{improper} PDF $\cc$). We provide a summarized pseudo-code of the proposed strategy in Algorithm~\ref{alg:RBDO}.\par

\begin{algorithm}[H]
  \caption{Adaptive surrogate-based nested RBDO strategy}
  \label{alg:RBDO}
  \begin{algorithmic}[1]
%     \STATE \COMMENT{Initializations}
    \STATE $\ve{\theta}^{(0)}$, $\ve{\theta}^{L}$, $\ve{\theta}^{U}$, $j := 0$
%     \STATE \COMMENT{Find bounds of the hyperrectangular confidence region}
    \FOR{$i=1$ \TO $n$}
      \STATE $q_{V_i}^- := \min\limits_{\ve{\theta}^{L} \leq \ve{\theta} \leq \ve{\theta}^{U}} F_{X_i}^{-1}\left(x_i,\,\ve{\theta}\right)$
      \STATE $q_{V_i}^+ := \max\limits_{\ve{\theta}^{L} \leq \ve{\theta} \leq \ve{\theta}^{U}} F_{X_i}^{-1}\left(x_i,\,\ve{\theta}\right)$
    \ENDFOR
    \STATE $w := \ve{v} \mapsto \fcar{\cd_{\ve{V}},\,\beta_0}{\ve{v}}$
    \STATE $k := \Phi^{-1}(97.5\%)$, $\varepsilon_{\beta} := 10^{-1}$
%     \STATE \COMMENT{Nested adaptive surrogated-based RBDO loop}
    \STATE Refine := \TRUE, Optimize := \TRUE
    \WHILE {Refine \AND Optimize}
%       \STATE \COMMENT{1. Build and refine a kriging model of the performance function in the hyperrectangular confidence region}
      \WHILE {Refine}
	\STATE $\widehat{G} := {\rm RefineKrigingModel}(w,\,k,\,\varepsilon_{\beta})$ \COMMENT{Use Algorithm~\ref{alg:DOE}}
      \ENDWHILE
%       \STATE \COMMENT{2. Perform reliability and reliability sensitivity analyses onto the surrogates}
      \STATE $\widehat{F}^{~0} := \acc{\ve{x}~:~\mu_{\widehat{G}}(\ve{x}) \leq 0}$
      \STATE $\widehat{\beta}^{(j)},\,\nabla_{\ve{\theta}}\widehat{\beta}^{(j)} := {\rm ReliabilityAnalysis}(\widehat{F}^{~0},\,\ve{\theta}^{(j)})$
%       \STATE \COMMENT{3. Compute cost, soft constrains and their gradients}
      \STATE $c^{(j)} := c(\ve{\theta}^{(j)})$, $\nabla_{\ve{\theta}}c^{(j)} := \nabla_{\ve{\theta}}c(\ve{\theta}^{(j)})$
      \STATE $\ve{f}^{(j)} := \ve{f}(\ve{\theta}^{(j)})$, $\nabla_{\ve{\theta}}\ve{f}^{(j)} := \nabla_{\ve{\theta}}\ve{f}(\ve{\theta}^{(j)})$
%       \STATE \COMMENT{4. Find optimization direction}
      \STATE $\ve{d}^{(j)} := {\rm SolveQuasiSQP}(c^{(j)},\,\ve{f}^{(j)},\,\widehat{\beta}^{(j)},\,\nabla_{\ve{\theta}}c^{(j)},\,\nabla_{\ve{\theta}}\ve{f}^{(j)},\,\nabla_{\ve{\theta}}\widehat{\beta}^{(j)})$
%       \STATE \COMMENT{5. Find optimal step size}
      \STATE $s^{(j)} := {\rm GoldsteinArmijoStepSizeRule}(c,\,\ve{f},\,\widehat{\beta},\,\ve{d}^{(j)})$
%       \STATE \COMMENT{6. Improve design}
      \STATE $\ve{\theta}^{(j+1)} := \ve{\theta}^{(j)} + s^{(j)}\,\ve{d}^{(j)}$
%       \STATE \COMMENT{7. Check the kriging model accuracy at the new design}
      \FOR {$i := -1,\,0,\,+1$}
	\STATE $\widehat{F}^{~i} := \acc{\ve{x}~:~\mu_{\widehat{G}}(\ve{x}) + i\,k\,\sigma_{\widehat{G}}(\ve{x}) \leq 0}$
	\STATE $\widehat{\beta}^{~i} := {\rm ReliabilityAnalysis}(\widehat{F}^{~i},\,\ve{\theta}^{(j+1)})$
      \ENDFOR
%       \STATE \COMMENT{8. Refine or done?}
      \STATE ${\rm Refine} := \max\left(\widehat{\beta}^{+1} - \widehat{\beta}^{~0},\,\widehat{\beta}^{~0} - \widehat{\beta}^{-1}\right) > \varepsilon_{\beta}$
%       \STATE \COMMENT{8. Optimize or done?}
      \STATE Optimize := $\left\|\ve{\theta}^{(j+1)} - \ve{\theta}^{(j)}\right\| > \varepsilon_{\ve{\theta}}$ \OR $\left\|c^{(j+1)} - c^{(j)}\right\| > \varepsilon_c$ \OR $\exists~i~\left|~f_i > 0\right.$ \OR $\widehat{\beta} < \beta_0$
    \ENDWHILE
  \end{algorithmic}
\end{algorithm}

The first step of the algorithm consists in finding the hyperrectangular region that bounds the confidence region of the augmented probability density function according to Section~\ref{sec:AugmentedReliabilitySpace}. Once this is done, one may define the uniform weighting density \added{$w\equiv\mathbbm{1}_{\cd_{\ve{V}},\,\beta_0}$ in \eqref{eq:AugmentedWeightingPDF}} and use it within the adaptive population-based refinement procedure detailed in Section~\ref{sec:DOE} and Algorithm~\ref{alg:DOE}. Kriging models are built for each performance function, and they are refined until they meet the selected accuracy regarding reliability estimation. \added{It is worth noting that the kriging surrogates are built in the augmented reliability space spanned by $\ve{V}$, but used in the current space of random variables spanned by $\ve{X}\mid\ve{\Theta}=\ve{\theta}^{(j)}$.} As soon as they are accurate enough we perform surrogate-based reliability and reliability sensitivity analysis in order to propose an improved design. A quasi-SQP algorithm is then used in order to determine the best improvement direction; and the Goldstein-Armijo approximate line-search rule is used to find the best step size along that direction. The current design is improved and the kriging model accuracy for reliability estimation is being checked at the improved design. The convergence is obtained if the optimization has converged (using the regular criteria in gradient-based deterministic optimization) and if the kriging models allow a sufficiently accurate reliability estimation according to the proposed error measure.\par

\section{\label{sec:Appli}Applications}

In this section, the proposed adaptive nested surrogate-based RBDO strategy is applied to some examples from the literature for performance comparison purposes. All the kriging surrogates are sequentially refined in order to achieve an empirical error measure $\varepsilon_{\beta} \leq 10^{-1}$ on the estimation of the reliability indices.\par

\subsection{Elastic buckling of a straight column -- An analytical reference}

In essence, the purpose of this first basic example is to validate the proposed algorithm with respect to a reference analytical solution.\par

Let us consider a long simply-supported rectangular column with section $b \times h$ subjected to a constant service axial load $F_{\rm ser}$. Provided $h\,\leq\,b$ and its constitutive material is characterized by a linear elastic behavior through its Young's modulus $E$, its critical buckling load is given by the Euler formula:
\begin{equation}
  F_{\rm cr} = \frac{\pi^2\,E\,b\,h^3}{12\,L^2}.
\end{equation}%
This allows one to formulate the performance function which will be involved in the probabilistic constraint as:%
\begin{equation}
  g\left(E,\,b,\,h,\,L\right) = \frac{\pi^2\,E\,b\,h^3}{12\,L^2} - F_{\rm ser}.
\end{equation}

The probabilistic model consists in the 3 independent random variables given in Table~\ref{tab:EulerBuckling_stoch}.

\begin{table}[b]
  \begin{footnotesize}
    \begin{center}
      \begin{tabular}{llcc}
	\hlineT \rowcolor{gray!30}
	\textbf{Variable} & \textbf{Distribution} & \textbf{Mean} & \textbf{C.o.V.} \\
	\hlineT
	$E$ (MPa) & Lognormal & $10\,000$ & $15\%$ \\
	$b$ (mm) & Lognormal & $\mu_b$ & $5\%$ \\
	$h$ (mm) & Lognormal & $\mu_h$ & $5\%$ \\
	$L$ (mm) & Deterministic & $3\,000$ & -- \\
	\hlineB
      \end{tabular}
    \end{center}
  \end{footnotesize}
  \caption{Elastic buckling of a straight column -- Probabilistic model}
  \label{tab:EulerBuckling_stoch}
\end{table}

Applying the performance function as a design rule in a fully deterministic fashion (\ie using the means of the random variates) allows to determine the service load $F_{\rm ser}$ so that the initial deterministic design $\mu_b^{(0)} = \mu_h^{(0)} = 200$~mm satisfies the limit-state equation $g(\ve{x})=0$:
\begin{equation}
  F_{\rm ser} = \frac{\pi^2\,\mu_E\,I_0}{L^2} = \frac{\pi^2\,\mu_E\,\mu_b^{(0)}\,\mu_h^{(0)\,3}}{12\,L^2}.
\end{equation}

The reliability-based design problem consists in finding the optimal means $\mu_b$ and $\mu_h$ of the random width $b$ and height $h$. The optimal design is the one that minimizes the average cross section area which is approximated as follows:%
\begin{equation}
 c(\ve{\theta}) = \mu_b\,\mu_h.
\end{equation}%
It should also satisfy the following deterministic constraint:%
\begin{equation}
 f\left(\ve{\theta}\right) = \mu_h - \mu_b \leq 0
\end{equation}%
in order to ensure that the Euler formula is applicable, as well as the following safety probabilistic constraint:%
\begin{equation}
 - \Phi^{-1}\left( \Prob{g\left(\ve{X}\right)\leq0} \right) \geq \beta_0
\end{equation}%
where $\beta_0 = 3$ is the generalized target reliability index.

\subsubsection{Analytical solution}

Due to the use of lognormal random variates in the probabilistic model together with the simple performance function (multiplications and divisions), the problem can be handled analytically. Indeed, the isoprobabilistic transform of the limit-state surface equation in terms of standard normal random variates is straightforward and leads to a linear equation. And it finally turns out after basic algebra that the Hasofer-Lind reliability index (associated with the exact failure probability) reads:%
\begin{equation}
 \beta_{\rm HL} = \frac{\log\left(\frac{\pi^2}{12\,F_{\rm ser}}\right) + \lambda_E + \lambda_b + 3\,\lambda_h - 2\,\lambda_L}{\sqrt{\zeta_E^2 + \zeta_b^2 + 9 \zeta_h^2 + 4 \zeta_L^2}}
\end{equation}%
where $\lambda_{\bullet}$ and $\zeta_{\bullet}$ denote the parameters of the lognormal random variates.\par

The optimal solution of the RBDO problem is then simply derived by saturating the two constraints in log-scale (\ie with respect to $\lambda_b$ and $\lambda_h$) and this leads to the square cross section with parameters:
\begin{equation}
  \lambda_b^* = \lambda_h^* = \frac{1}{4}\left(\beta_0\,\sqrt{\zeta_E^2 + \zeta_b^2 + 9 \zeta_h^2 + 4 \zeta_L^2} - \log\left(\frac{\pi^2}{12\,F_{\rm ser}}\right) - \lambda_E + 2\,\lambda_L\right).
\end{equation}

\subsubsection{Numerical solution}

The proposed numerical strategy is applied in order to solve the RBDO problem numerically. The refinement procedure of the limit-state surface is initialized with an initial DOE of $K_0=10$ points and $K=10$ points are sequentially added to the DOE if it is not accurate enough for reliability estimation.\par

\begin{figure}[!h]
  \begin{center}
    \subfigure[Run \#1 -- Starting from the optimal deterministic design]{\includegraphics[width=.8\textwidth]{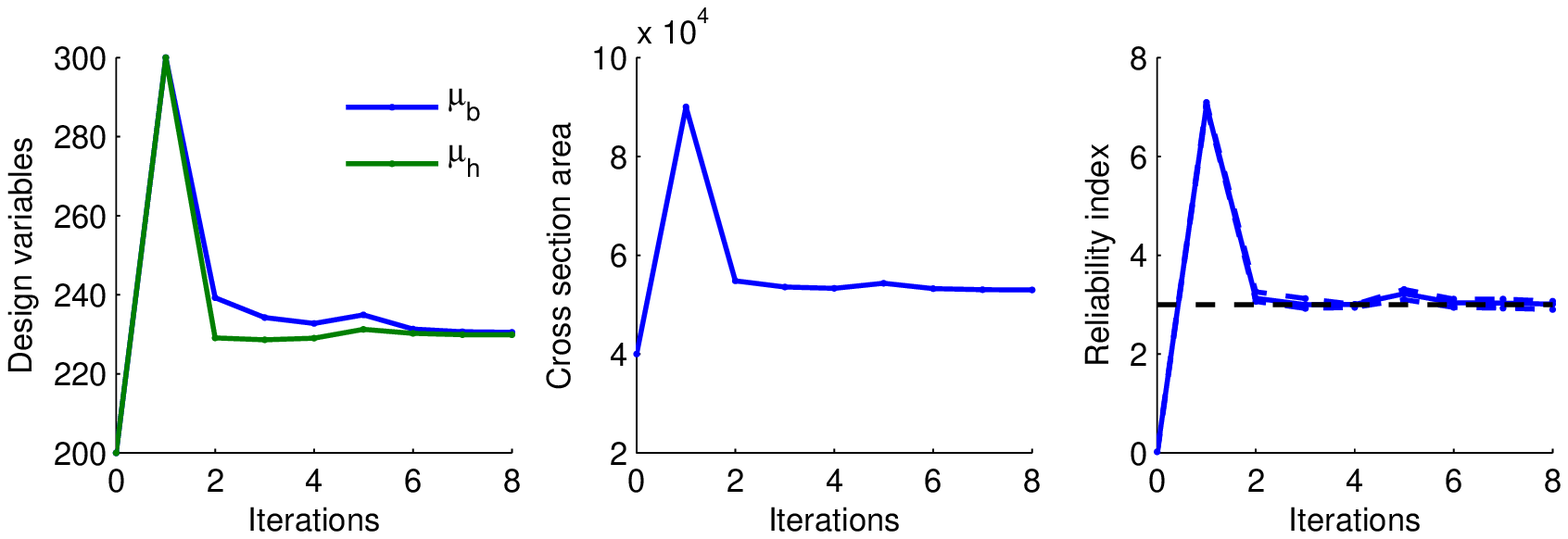}}\\
    \subfigure[Run \#2 -- Starting from a conservative design]{\includegraphics[width=.8\textwidth]{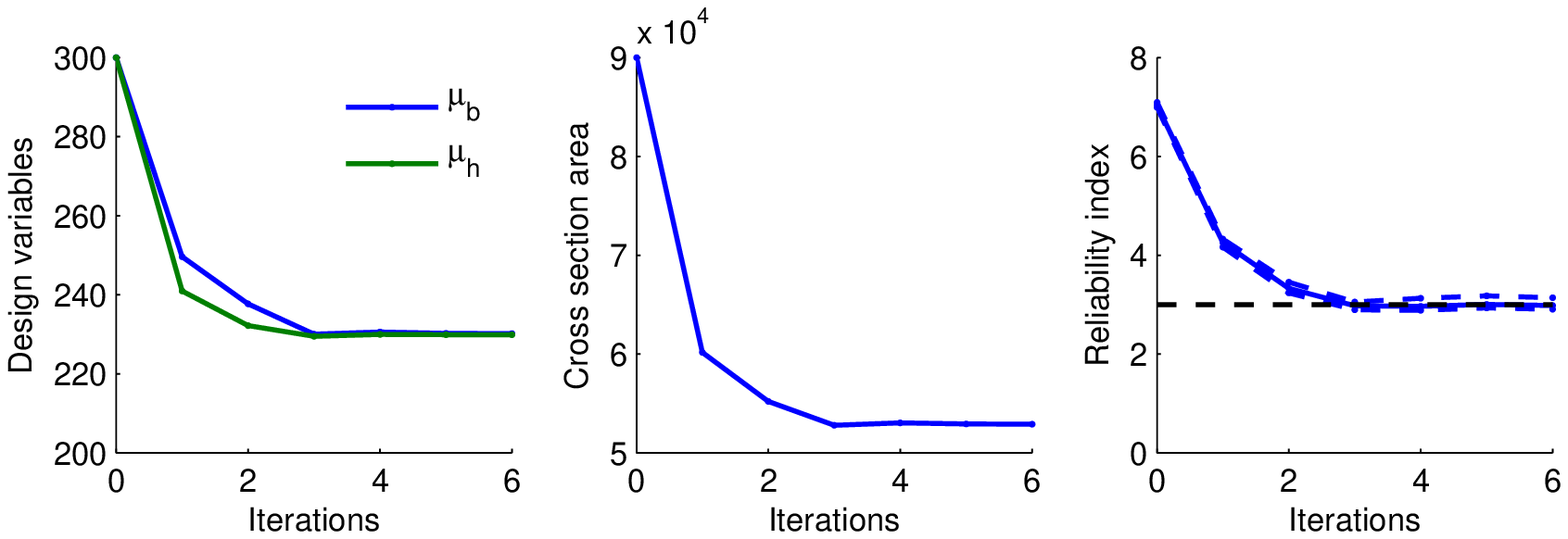}}
  \end{center}
  \caption{Elastic buckling of a straight column -- Convergence}
  \label{fig:EulerBuckling_convergence}
\end{figure}

The convergence of the algorithm is depicted in Figure~\ref{fig:EulerBuckling_convergence} for two runs starting from different initial designs. Run \#1 is initiated with the optimal deterministic design $\mu_b = \mu_h = b^* = h^* = 200$~mm whereas Run \#2 is initiated with an oversized design $\mu_b = \mu_h = 300$~mm. Convergence is achieved as all the constraints (deterministic and reliability-based) are satisfied and both the cost and design variables have reached a stable value. The algorithm converges to the exact solution derived in the previous subsection which is the square section with width $\mu_b^* = \mu_h^* \approx 231$~mm -- the approximation of the exact solution is only due to the numerical error. Note that the reliability-based optimal design is 15\% higher than the optimal deterministic design for the chosen reliability level ($\beta_0 = 3$).\par

This optimum is reached using only 20 evaluations of the performance function thanks to the kriging surrogate. The DOE used for this purpose is enriched only once and it is then accurate enough for all the design configurations including the optimal design. Running the same RBDO algorithm without using the kriging surrogates (\ie using subset simulation onto the real performance function for the nested reliability and reliability sensitivity analyses) requires about $4\times10^6$ evaluations of the performance function for the same number of iterations of the optimizer and converges to the same optimal design.\par

\subsection{Short column case}

This simple mechanical example is extensively used in the RBDO literature as a benchmark for numerical methods in RBDO. In this paper, we use the results from the article by \citet{Royset2001} as reference. It consists in a short column with rectangular cross section $b \times h$. It is subjected to an axial load $F$ and two bending moments $M_1$ and $M_2$ whose axes are defined with respect to the two axes of inertia of the cross section. Such a load combination is referred to as \emph{oblique bending} due to the rotation of the neutral axis. Under the additional assumption that the material is elastic perfectly plastic, the performance function describing the ultimate serviceability of the column with respect to its yield stress $f_y$ reads as follows:%
\begin{equation} \label{eq:Column_lsf}
  G\left(M_1,\,M_2,\,P,\,f_y,\,b,\,h\right) = 1 - \frac{4\,M_1}{b\,h^2\,f_y} - \frac{4\,M_2}{b^2\,h\,f_y} - \left(\frac{F}{b\,h\,f_y}\right)^2.
\end{equation}

The stochastic model originally involves three independent random variables whose distributions are given in Table~\ref{tab:Column_StoModel}. Note that in the original paper, the design variables $\mu_b$ and $\mu_h$ are considered as deterministic. Since the present approach only deals with design variables that defines the joint PDF of the random vector $\ve{X}$, they are considered here as Gaussian with a small coefficient of variation and the optimization is performed with respect to their mean $\mu_b$ and $\mu_h$. \removed{And the} \added{The} objective function is formulated as follows:
\begin{equation} \label{eq:RBDO_Royset}
 c(\mu_b,\,\mu_h) = c_0(\mu_b,\,\mu_h) + P_f(\mu_b,\,\mu_h)\,c_f(\mu_b,\,\mu_h) = \mu_b\,\mu_h\,(1+100\,P_f)
\end{equation}
where the product $P_f \times c_f$ is the expected failure cost which is chosen as proportional to the construction cost $c_0$. The search for the optimal design is limited to the designs that satisfy the following geometrical constraints: $1/2 \leq \mu_b/\mu_h \leq 2$ with $100 \leq \mu_b,\,\mu_h \leq 1000$, and the minimum reliability is chosen as $\beta_0=3$.

\begin{table}[!h]
 \begin{footnotesize}
  \begin{center}
    \begin{tabular}{lllcc}
      \hlineT \rowcolor{gray!30}
      \multicolumn{2}{l}{\cellcolor{gray!30}\textbf{Variable}} & \textbf{Distribution} & \textbf{Mean} & \textbf{C.o.V.} \\
      \hlineT
      $M_1$ & (N.mm)      & Lognormal             &$250\times10^6$&       30\%      \\
      $M_2$ & (N.mm)      & Lognormal             &$125\times10^6$&       30\%      \\
      $P$ & (N)           & Lognormal             &$2.5\times10^6$&       20\%      \\
      $f_y$ & (MPa)       & Lognormal             &      40       &       10\%      \\
      $b$ & (mm)          & Gaussian              &   $\mu_b$     &        1\%$^*$  \\
      $h$ & (mm)          & Gaussian              &   $\mu_h$     &        1\%$^*$  \\
    \hlineB
    \multicolumn{5}{l}{\footnotesize{$^*$ Additional uncertainty introduced to match the RBDO problem formulation in \eqref{eq:RBDO_RIA}}}
   \end{tabular}
  \end{center}
 \end{footnotesize}
 \caption{Short column case from \citet{Royset2001} -- Probabilistic model}
 \label{tab:Column_StoModel}
\end{table}

The results are given in Table~\ref{tab:Column_results}. In this table, $\beta_{\rm HL}$ denotes the Hasofer-Lind reliability index (FORM-based), and $\beta_{\rm sim}$ denotes the generalized reliability index estimated by \emph{subset simulation} with a coefficient of variation less than 5\%. The deterministic design optimization (DDO) was performed using the mean values of all the variables in Table~\ref{tab:Column_StoModel} without considering uncertainty and thus leads to a 50\% failure probability. Note that the corresponding optimal cost does not account for the expected failure cost. The other lines of Table~\ref{tab:Column_results} shows the results of the RBDO problem. The first row gives the reference results from \citet{Royset2001}. The number of performance function calls was not given in the original paper. However it may be estimated to $4\times10^6$ given the methodology the authors used and assuming they targeted a 5\% coefficient of variation on the failure probability in their Monte-Carlo simulation. The second row provides the results from a FORM-based nested RBDO algorithm (RIA). This latter approach seems to lead to a slightly better design though it is due to the first-order reliability assumptions that are not conservative in this case. Indeed, subset simulation leads to a little lower generalized reliability index $\beta_{\rm sim} \approx 3.19$ (with a 5\% C.o.V.) which in turns slightly increases the failure-dependent objective function to $c \approx 2.20\times10^5~{\rm mm}^2$. This example shows that FORM-based approaches can mistakenly lead to non conservative optimal designs -- without any self-quantification of the possible degree of non conservatism. The third row gives the results obtained by the same nested RBDO algorithm, using however the subset simulation technique as the reliability (and reliability sensitivity) estimator. Finally, the last row gives the results obtained when using kriging as a surrogate for the limit-state surface. The kriging model used for this application used a constant regression model and a squared exponential autocovariance model. It was initialized with a 50-point DOE and sequentially refined with $K=10$ points per refinement iteration.\par

\begin{table}[t]
  \begin{footnotesize}
    \begin{center}
      \begin{tabular}{llccr}
	\hlineT \rowcolor{gray!30}
	\textbf{Method}       & \textbf{Opt. design} (mm)                        & \textbf{Cost} (mm$^2$) & \textbf{\# of func. calls} & \textbf{Reliability} \\
	\hlineB
	\textbf{DDO}          & $\begin{array}{cc}b = 258 & h = 500 \end{array}$ &   $1.29\times10^5$  &             50                & $\beta_{\rm sim}\approx0.01$\\
	\hlineB
	\multicolumn{5}{l}{\textbf{RBDO ($\beta \geq 3$)}}\\
	Reference (DSA)       & $\begin{array}{cc}b = 372 & h = 559 \end{array}$ &   $2.15\times10^5$   &            --                & $\beta_{\rm sim}\approx3.38$\\
	FORM-based (RIA)      & $\begin{array}{cc}b = 399 & h = 513 \end{array}$ &   $2.12\times10^5$   &         9\,472               & $\beta_{\rm HL}=3.38$\\
% 	\multicolumn{2}{r}{\textit{Check of FORM assumptions}}                   &   $2.20\times10^5$   &        $4\times10^5$         & $\beta_{\rm sim}\approx3.19$\\ % IN THE TEXT !!!
	Present w/o kriging   & $\begin{array}{cc}b = 369 & h = 560 \end{array}$ &   $2.16\times10^5$   &        $19\times10^6$        & $\beta_{\rm sim}\approx3.35$\\
	Present w/ kriging    & $\begin{array}{cc}b = 379 & h = 547 \end{array}$ &   $2.17\times10^5$   &            140               & $\beta_{\rm sim}\approx3.32$\\
	\hlineT
      \end{tabular}
    \end{center}
  \end{footnotesize}
  \caption{Short column case from \citet{Royset2001} -- Comparative results}
  \label{tab:Column_results}
\end{table}

\begin{figure}[H]
  \begin{center}
    \includegraphics[width=.8\textwidth]{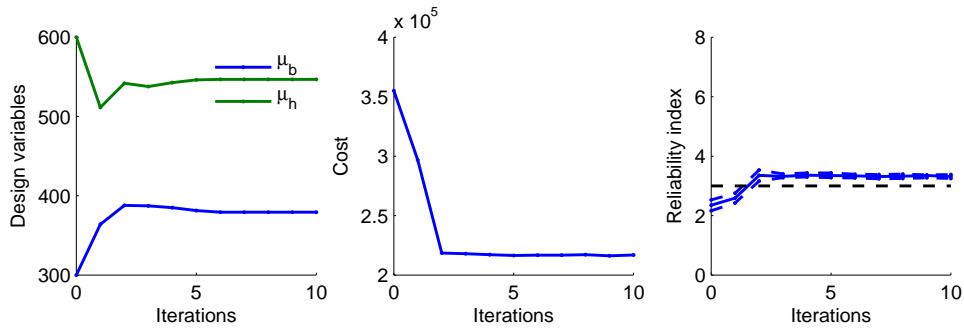}
  \end{center}
  \caption{Short column case -- Convergence}
  \label{fig:Column_convergence}
\end{figure}

Another interesting fact about this example is that the reliability constraint is not saturated at the optimum: the algorithm converges at a higher reliability level as illustrated in Figure~\ref{fig:Column_convergence}. This is due to the specific formulation of the cost function in \eqref{eq:RBDO_Royset} that accounts for a failure cost that is indexed onto the failure probability. Indeed, the cost function behaves itself as a constraint and the optimal reliability level is formulated in terms of an acceptable risk (probability of occurrence times consequence) instead of an acceptable reliability index $\beta_0$.\par

\subsection{Bracket structure case}

\begin{figure}
  \begin{center}
    \includegraphics[width=.8\textwidth]{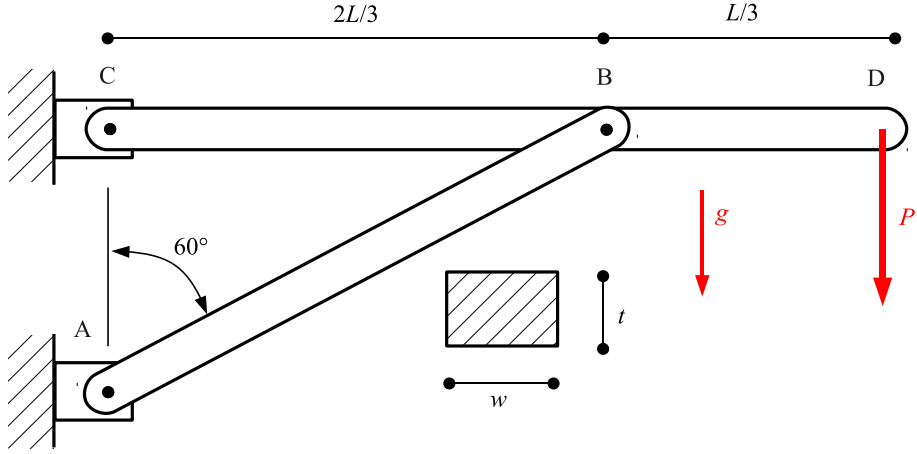}
  \end{center}
  \caption{Bracket structure case -- Mechanical model}
  \label{fig:Bracket_mech}
\end{figure}

This mechanical example is originally taken from \citet{Chateauneuf2008}. It consists in the study of the failure modes of the bracket structure pictured in Figure~\ref{fig:Bracket_mech}. The bracket structure is loaded by its own weight due to gravity and by an additional load at the right tip. The two failure modes under considerations are:%
\begin{itemize}
  \item the maximum bending in the horizontal beam (CD, at point B) should not exceed the yield strength of the constitutive material, so that the first performance function reads as follows:%
  \begin{equation} \label{eq:Bracket_lsf1}
    g_1\left(w_{\rm CD},\,t,\,L,\,P,\,\rho,\,f_y\right) = f_y - \sigma_{\rm B}
  \end{equation}%
  where the maximum bending stress reads:
  \begin{equation}
    \sigma_{\rm B} = \frac{6\,M_{\rm B}}{w_{\rm CD}\,t^2} \quad \text{with:} \quad M_{\rm B} = \frac{P\,L}{3} + \frac{\rho\,g\,w_{\rm CD}\,t\,L^2}{18},
  \end{equation}
  \item the maximum axial load in the inclined member (AB) should not exceed the Euler critical buckling load (neglecting its own weight), so that the second performance function reads as follows:%
  \begin{equation} \label{eq:Bracket_lsf2}
    g_2\left(w_{\rm AB},\,w_{\rm CD},\,t,\,L,\,P,\,\rho,\,f_y\right) = F_{\rm buckling} - F_{\rm AB}
  \end{equation}
  where the critical Euler buckling load $F_{\rm buckling}$ is defined as:
  \begin{equation}
    F_{\rm buckling} = \frac{\pi^2\,E\,I}{L_{\rm AB}^2} = \frac{\pi^2\,E\,t\,w_{\rm AB}^3\,9\,\sin^2\theta}{48\,L^2},
  \end{equation}
  and the resultant of axial forces in member AB reads (neglecting its own weight):
  \begin{equation}
    F_{\rm AB} = \frac{1}{\cos\theta}\,\left(\frac{3\,P}{2} + \frac{3\,\rho\,g\,w_{\rm CD}\,t\,L}{4}\right).
  \end{equation}
\end{itemize}
The probabilistic model for this example is the collection of independent random variables given in Table~\ref{tab:Bracket_stoch}. Note that the coefficient of variation of the random design variables is kept constant along the optimization as in the original paper.

\begin{table}[H]
  \begin{footnotesize}
    \begin{center}
      \begin{tabular}{lllcc}
	\hlineT \rowcolor{gray!30}
	\multicolumn{2}{l}{\cellcolor{gray!30}\textbf{Variable}} & \textbf{Distribution} & \textbf{Mean} & \textbf{C.o.V.} \\
	\hlineT
	$P$ & (kN)          & Gumbel                &      100      &       15\%      \\
	$E$ & (GPa)         & Gumbel                &      200      &        8\%      \\
	$f_y$ & (MPa)       & Lognormal             &      225      &        8\%      \\
	$\rho$ & (kg/m$^3$) & Weibull               &     7\,860    &       10\%      \\
	$L$ & (m)           & Gaussian              &       5       &        5\%      \\
	$w_{\rm AB}$ & (mm) & Gaussian              & $\mu_{w_{\rm AB}}$  &        5\%      \\
	$w_{\rm CD}$ & (mm) & Gaussian              & $\mu_{w_{\rm CD}}$  &        5\%      \\
	$t$ & (mm)          & Gaussian              &      $\mu_t$        &        5\%      \\
	\hlineT
      \end{tabular}
    \end{center}
  \end{footnotesize}
  \caption{Bracket structure case -- Probabilistic model}
  \label{tab:Bracket_stoch}
\end{table}

The RBDO problem consists in finding the rectangular cross sections of the two structural members that minimize the expected overall structural weight which is approximated as follows:
\begin{equation}
  c\left(\mu_{w_{\rm AB}},\,\mu_{w_{\rm CD}},\,t\right) = \mu_\rho\,\mu_t\,\mu_L\,\left(\frac{4\,\sqrt{3}}{9}\,\mu_{w_{\rm AB}} + \mu_{w_{\rm CD}}\right)
\end{equation}
while satisfying a minimum reliability requirement equal to $\beta_0=2$ with respect to the two limit-states in \eqref{eq:Bracket_lsf1} and \eqref{eq:Bracket_lsf2} considered independently. The search for the optimal design is bounded to the following hyperrectangle: $50 \leq \mu_{w_{\rm AB}},\,\mu_{w_{\rm CD}},\,\mu_t \leq 300$ (in mm).\par

The comparative results are provided in Table~\ref{tab:Bracket_results} considering the results from \citet{Chateauneuf2008} as reference.\par

\begin{table}[H]
  \begin{footnotesize}
    \begin{center}
      \begin{tabular}{llccc}
	\hlineT \rowcolor{gray!30}
	\textbf{Method}       & \textbf{Opt. design} (mm) & \textbf{Cost} (kg) & \textbf{\# of func. calls} & \textbf{Reliability} \\
	\hlineT
	\textbf{DDO w/ PSF}$^*$ & $\begin{array}{rcl}w_{\rm AB} & = & 61 \\ w_{\rm CD} & = & 202 \\ t & = & 269 \end{array}$ &   2\,632  &             40                & $\begin{array}{rcl}\beta_{\rm sim\,1} & \approx & 4.84 \\ \beta_{\rm sim\,2} & \approx & 2.82\end{array}$ \\
	\hlineB
	\multicolumn{5}{l}{\textbf{RBDO ($\beta_i \geq 2, i=1,\,2$)}}\\
	SORA$^*$      & $\begin{array}{rcl}w_{\rm AB} & = & 61 \\ w_{\rm CD} & = & 157 \\ t & = & 209 \end{array}$ &   1\,675  &           1\,340              & $\begin{array}{rcl}\beta_{\rm sim\,1} & \approx & 1.96 \\ \beta_{\rm sim\,2} & \approx & 2.01\end{array}$ \\
	\hline
	RIA$^*$       & $\begin{array}{rcl}w_{\rm AB} & = & 61 \\ w_{\rm CD} & = & 157 \\ t & = & 209 \end{array}$ &   1\,675  &           2\,340              & $\begin{array}{rcl}\beta_{\rm sim\,1} & \approx & 1.96 \\ \beta_{\rm sim\,2} & \approx & 2.01\end{array}$ \\
	\hline
	Present w/o kriging   & $\begin{array}{rcl}w_{\rm AB} & = & 58 \\ w_{\rm CD} & = & 119 \\ t & = & 241 \end{array}$   &   1\,550  &       $\begin{array}{c} 5\times10^6 \\ 5\times10^6\end{array}$ & $\begin{array}{rcl}\beta_{\rm sim\,1} & \approx & 2.00 \\ \beta_{\rm sim\,2} & \approx & 2.01\end{array}$ \\
	\hline
	Present w/ kriging    & $\begin{array}{rcl}w_{\rm AB} & = & 59 \\ w_{\rm CD} & = & 135 \\ t & = & 226 \end{array}$ &   1\,610  &       $\begin{array}{c} 100 \\ 150\end{array}$ & $\begin{array}{rcl}\beta_{\rm sim\,1} & \approx & 2.01 \\ \beta_{\rm sim\,2} & \approx & 2.03\end{array}$ \\
	\hlineT
	\multicolumn{5}{l}{$^*$ As computed by \citet{Chateauneuf2008}.}
      \end{tabular}
    \end{center}
  \end{footnotesize}
  \caption{Bracket structure case -- Comparative results}
  \label{tab:Bracket_results}
\end{table}

The first row gives the deterministic optimal design that was obtained by \citet{Chateauneuf2008} using \emph{partial safety factors} (PSF). It can be seen from the reliability indices that these PSF provide a significant safety level. However, one may rather want to find an even lighter design allowing for a lower safety level $\beta_0 = 2$. To do so, the RBDO formulation of the problem is solved. \citet{Chateauneuf2008} used the SORA technique which is a decoupled FORM-based approach. The reliability indices at the optimal design were checked using the subset simulation technique (targeting a coefficient of variation less than 5\%) and revealed that the FORM-based approach slightly underestimates the first optimal reliability index in this case. The RIA technique which is a standard double-loop FORM-based approach provides the same solution but it is less efficient.\par

\begin{figure}[H]
  \begin{center}
    \includegraphics[width=\textwidth]{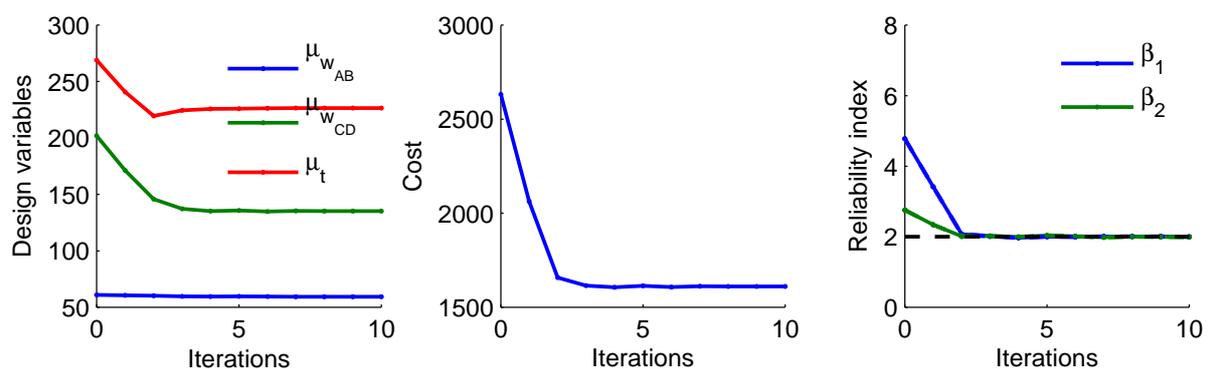}
  \end{center}
  \caption{Bracket structure case -- Convergence}
  \label{fig:Bracket_convergence}
\end{figure}

Implementing the proposed approach without plugging the kriging surrogates converges to similar results but clearly confirms that direct simulation-based approaches are not tractable for RBDO. Replacing the performance function by their kriging counterparts allows to save a significant number of simulations ($10^2$ opposed to $10^6$) and in addition, to provide an error measure on the reliability estimation as opposed to the FORM-based approaches. However, one may note the disparities between the proposed designs. First, the disparity between FORM-based methods and the presently proposed strategies is explained by the conservatism of the FORM assumptions in this case. Then, the disparity between the two present approaches is certainly due to the flatness of the sub-optimization problem and the stochastic nature of the simulation-based reliability estimation. The convergence of the algorithm is depicted in Figure~\ref{fig:Bracket_convergence}.\par

\newpage

\section{\label{sec:Discussion}Conclusion}

The aim of the present paper was to develop a strategy for solving reliability-based design optimization (RBDO) problems that is applicable to engineering problems involving time-consuming performance models. Starting with the premise that simulation-based approaches are not affordable when the performance function involves the output of an expensive-to-evaluate computer model, and that the MPFP-based approaches do not allow to quantify the error on the estimation of the failure probability, an approach based on kriging and subset simulation is explored.\par

The strategy has been tested on a set of examples from the RBDO literature and proved to be competitive with respect to its FORM-based counterparts. Indeed, convergence is achieved with only a few dozen evaluations of the real performance functions. In contrast with the FORM-based approaches, the proposed error measure allows one to quantify and sequentially minimize the surrogate error onto the final quantity of interest: the optimal failure probability. It is important to note that the numerical efficiency of the proposed strategy mainly relies upon the properties of the space where the kriging surrogates are built: the so-called \emph{augmented reliability space}. This space is obtained by considering that the design variables in the RBDO problem simply augments the uncertainty in the random vector involved in the reliability problem. Building the surrogates in such a space allows one to reuse them from one RBDO iteration to the other and thus saves a large number of performance functions evaluations. \added{It is also worth noting that the original refinement strategy proposed in Section \ref{sec:DOE} makes it possible to add several observations in the design of experiments at the same time, and thus to benefit from the availability of a distributed computing platform to speed up convergence.}\par

However, as already mentioned in the literature, it was observed that the number of experiments increases with the number of variables involved in the performance functions and that the kriging strategy loses numerical efficiency when the DOE contains more than a few thousands experiments -- although such an amount of information is not even available in real-world engineering cases. This latter point requires further investigation. A problem involving a nonlinear-finite-element-based performance function and 10 variables is currently investigated and will be published in a forthcoming paper.\par

\begin{acknowledgements}
The first author was funded by a CIFRE grant from Phimeca Engineering S.A. subsidized by the ANRT (convention number 706/2008). The financial support from DCNS is also gratefully acknowledged.
\end{acknowledgements}

\newpage

% BibTeX users please use one of
\bibliographystyle{spbasic}      % basic style, author-year citations
%\bibliographystyle{spmpsci}      % mathematics and physical sciences
%\bibliographystyle{spphys}       % APS-like style for physics
% \bibliography{biblioEMP,phimecabiblio}   % name your BibTeX data base

\begin{thebibliography}{54}
\providecommand{\natexlab}[1]{#1}
\providecommand{\url}[1]{{#1}}
\providecommand{\urlprefix}{URL }
\expandafter\ifx\csname urlstyle\endcsname\relax
  \providecommand{\doi}[1]{DOI~\discretionary{}{}{}#1}\else
  \providecommand{\doi}{DOI~\discretionary{}{}{}\begingroup
  \urlstyle{rm}\Url}\fi
\providecommand{\eprint}[2][]{\url{#2}}

\bibitem[{Aoues and Chateauneuf(2010)}]{Aoues2010}
Aoues Y, Chateauneuf A (2010) Benchmark study of numerical methods for
  reliability-based design optimization. Struct Multidisc Optim 41(2):277--294

\bibitem[{Au and Beck(2001)}]{Au2001}
Au S, Beck J (2001) Estimation of small failure probabilities in high
  dimensions by subset simulation. Prob Eng Mech 16(4):263--277

\bibitem[{Au(2005)}]{Au2005}
Au SK (2005) {Reliability-based design sensitivity by efficient simulation}.
  Computers \& Structures 83(14):1048--1061

\bibitem[{Berveiller et~al.(2006)Berveiller, Sudret, and
  Lemaire}]{Berveiller2006a}
Berveiller M, Sudret B, Lemaire M (2006) Stochastic finite elements: a non
  intrusive approach by regression. Eur J Comput Mech 15(1-3):81--92

\bibitem[{Bichon et~al.(2008)Bichon, Eldred, Swiler, Mahadevan, and
  McFarland}]{Bichon2008}
Bichon B, Eldred M, Swiler L, Mahadevan S, McFarland J (2008) {Efficient global
  reliability analysis for nonlinear implicit performance functions}. AIAA
  Journal 46(10):2459--2468

\bibitem[{Blatman and Sudret(2008)}]{BlatmanCras2008}
Blatman G, Sudret B (2008) Sparse polynomial chaos expansions and adaptive
  stochastic finite elements using a regression approach. Comptes Rendus
  M\'{e}canique 336(6):518--523

\bibitem[{Blatman and Sudret(2010)}]{BlatmanPEM2010}
Blatman G, Sudret B (2010) An adaptive algorithm to build up sparse polynomial
  chaos expansions for stochastic finite element analysis. Prob Eng Mech
  25(2):183--197

\bibitem[{Bourinet et~al.(2009)Bourinet, Mattrand, and Dubourg}]{Bourinet2009}
Bourinet JM, Mattrand C, Dubourg V (2009) {A review of recent features and
  improvements added to FERUM software}. In: Proc. ICOSSAR'09, Int Conf. on
  Structural Safety And Reliability, Osaka, Japan

\bibitem[{Bourinet et~al.(2010)Bourinet, Deheeger, and Lemaire}]{Bourinet2010}
Bourinet JM, Deheeger F, Lemaire M (2010) Assessing small failure probabilities
  by combined subset simulation and support vector machines. Submitted to
  Structural Safety

\bibitem[{Bucher and Bourgund(1990)}]{Bucher1990}
Bucher C, Bourgund U (1990) {A fast and efficient response surface approach for
  structural reliability problems}. Structural Safety 7(1):57--66

\bibitem[{Chateauneuf and Aoues(2008)}]{Chateauneuf2008}
Chateauneuf A, Aoues Y (2008) Structural design optimization considering
  uncertainties, Taylor \& Francis, chap~9, pp 217--246

\bibitem[{Cressie(1993)}]{Cressie1993}
Cressie N (1993) {Statistics for spatial data, revised edition}. John Wiley \&
  Sons Inc.

\bibitem[{Das and Zheng(2000)}]{Das2000}
Das PK, Zheng Y (2000) {Cumulative formation of response surface and its use in
  reliability analysis}. Prob Eng Mech 15(4):309--315

\bibitem[{Deheeger(2008)}]{Deheeger2008}
Deheeger F (2008) {Couplage m\'ecano-fiabiliste, $^2$SMART m\'ethodologie
  d'apprentissage stochastique en fiabilit\'e}. PhD thesis, {Universit\'e
  Blaise Pascal - Clermont II}

\bibitem[{Deheeger and Lemaire(2007)}]{Deheeger2007}
Deheeger F, Lemaire M (2007) {Support vector machine for efficient subset
  simulations: 2SMART method}. In: Proc. 10th~Int. Conf. on Applications of
  Stat. and Prob. in Civil Engineering (ICASP10), Tokyo, Japan

\bibitem[{Der~Kiureghian and Ditlevsen(2009)}]{DerKiureghian2009}
Der~Kiureghian A, Ditlevsen O (2009) Aleatory or epistemic? does it matter?
  Structural Safety 31(2):105--112

\bibitem[{Ditlevsen and Madsen(1996)}]{Ditlevsen1996}
Ditlevsen O, Madsen H (1996) {Structural reliability methods}, {Internet
  (v2.3.7, June-Sept 2007)} edn. John Wiley \& Sons Ltd, Chichester

\bibitem[{Du and Chen(2004)}]{Du2004}
Du X, Chen W (2004) {Sequential optimization and reliability assessment method
  for efficient probabilistic design}. J Mech Des 126:225--233

\bibitem[{Enevoldsen and Sorensen(1994)}]{Enevoldsen1994}
Enevoldsen I, Sorensen JD (1994) {Reliability-based optimization in structural
  engineering}. Structural Safety 15(3):169--196

\bibitem[{Faravelli(1989)}]{Faravelli1989}
Faravelli L (1989) {Response surface approach for reliability analysis}. J Eng
  Mech 115(12):2763--2781

\bibitem[{Hurtado(2004)}]{Hurtado2004}
Hurtado J (2004) An examination of methods for approximating implicit limit
  state functions from the viewpoint of statistical learning theory. Structural
  Safety 26:271--293

\bibitem[{Jensen et~al.(2009)Jensen, Valdebenito, Schu\"eller, and
  Kusanovic}]{Jensen2009}
Jensen H, Valdebenito M, Schu\"eller G, Kusanovic D (2009) Reliability-based
  optimization of stochastic systems using line search. Comput Methods Appl
  Mech Engrg 198(49-52):3915--3924

\bibitem[{Jones et~al.(1998)Jones, Schonlau, and Welch}]{Jones1998}
Jones D, Schonlau M, Welch W (1998) {Efficient global optimization of expensive
  black-box functions}. J Global Optim 13(4):455--492

\bibitem[{Kharmanda et~al.(2002)Kharmanda, Mohamed, and Lemaire}]{Kharmanda2002}
Kharmanda G, Mohamed A, Lemaire M (2002) Efficient reliability-based design
  optimization using a hybrid space with application to finite element
  analysis. Struct Multidisc Optim 24(3):233--245

\bibitem[{Kim and Na(1997)}]{Kim1997}
Kim SH, Na SW (1997) {Response surface method using vector projected sampling
  points}. Structural Safety 19(1):3--19

\bibitem[{Kirjner-Neto et~al.(1998)Kirjner-Neto, Polak, and
  Der~Kiureghian}]{KirjnerNeto1998}
Kirjner-Neto C, Polak E, Der~Kiureghian A (1998) {An outer approximation
  approach to reliability-based optimal design of structures}. J Optim Theory
  Appl 98:1--16

\bibitem[{Kuschel and Rackwitz(1997)}]{Kuschel1997}
Kuschel N, Rackwitz R (1997) {Two basic problems in reliability-based
  structural optimization}. Math Meth Oper Research 46(3):309--333

\bibitem[{Lebrun and Dutfoy(2009)}]{Lebrun2009a}
Lebrun R, Dutfoy A (2009) An innovating analysis of the {N}ataf transformation
  from the copula viewpoint. Prob Eng Mech 24(3):312--320

\bibitem[{Lee and Jung(2008)}]{Lee2008}
Lee T, Jung J (2008) {A sampling technique enhancing accuracy and efficiency of
  metamodel-based RBDO: Constraint boundary sampling}. Computers \& Structures
  86(13-14):1463--1476

\bibitem[{Lophaven et~al.(2002)Lophaven, Nielsen, and
  S{\o}ndergaard}]{Lophaven2002}
Lophaven S, Nielsen H, S{\o}ndergaard J (2002) {DACE, A Matlab Kriging
  Toolbox}. Technical University of Denmark

\bibitem[{MacQueen(1967)}]{MacQueen1967}
MacQueen J (1967) Some methods for classification and analysis of multivariate
  observations. In: Le~Cam J LM \&~Neyman (ed) Proc. 5$^{th}$ Berkeley Symp. on
  Math. Stat. \& Prob., University of California Press, Berkeley, CA, vol~1, pp
  281--297

\bibitem[{Neal(2003)}]{Neal2003}
Neal R (2003) Slice sampling. Annals Stat 31:705--767

\bibitem[{Oakley(2004)}]{Oakley2004b}
Oakley J (2004) Estimating percentiles of uncertain computer code outputs. J
  Roy Statist Soc Ser C 53(1):83--93

\bibitem[{Papadrakakis and Lagaros(2002)}]{Papadrakakis2002}
Papadrakakis M, Lagaros N (2002) {Reliability-based structural optimization
  using neural networks and Monte Carlo simulation}. Comput Methods Appl Mech
  Engrg 191(32):3491--3507

\bibitem[{Picheny et~al.(2010)Picheny, Ginsbourger, Roustant, and
  Haftka}]{Picheny2010b}
Picheny V, Ginsbourger D, Roustant, Haftka R (2010) Adaptive designs of
  experiments for accurate approximation of a target region. J Mech Des 132(7)

\bibitem[{Polak(1997)}]{Polak1997}
Polak E (1997) Optimization algorithms and consistent approximations. Springer

\bibitem[{Ranjan et~al.(2008)Ranjan, Bingham, and Michailidis}]{Ranjan2008}
Ranjan P, Bingham D, Michailidis G (2008) Sequential experiment design for
  contour estimation from complex computer codes. Technometrics 50(4):527--541

\bibitem[{Rasmussen and Williams(2006)}]{Rasmussen2006}
Rasmussen C, Williams C (2006) Gaussian processes for machine learning,
  {Internet} edn. Adaptive computation and machine learning, MIT Press,
  Cambridge, Massachusetts

\bibitem[{Robert and Casella(2004)}]{Robert2004}
Robert C, Casella G (2004) {Monte Carlo statistical methods (2$^{nd}$ Ed.)}.
  Springer Series in Statistics, Springer Verlag

\bibitem[{Royset and Polak(2004{\natexlab{a}})}]{Royset2004}
Royset J, Polak E (2004{\natexlab{a}}) Reliability-based optimal design using
  sample average approximations. Prob Eng Mech 19:331--343

\bibitem[{Royset and Polak(2004{\natexlab{b}})}]{Royset2004a}
Royset JO, Polak E (2004{\natexlab{b}}) {Implementable algorithm for stochastic
  optimization using sample average approximations}. J Optim Theory Appl
  122(1):157--184

\bibitem[{Royset et~al.(2001)Royset, Der~Kiureghian, and Polak}]{Royset2001}
Royset JO, Der~Kiureghian A, Polak E (2001) Reliability-based optimal
  structural design by the decoupling approach. Reliab Eng Sys Safety
  73(3):213--221

\bibitem[{Santner et~al.(2003)Santner, Williams, and Notz}]{Santner2003}
Santner T, Williams B, Notz W (2003) {The design and analysis of computer
  experiments}. Springer series in Statistics, Springer

\bibitem[{Severini(2005)}]{Severini2005}
Severini T (2005) Elements of distribution theory. Cambridge series in
  statistical and probabilistic mathematics, Cambridge University Press

\bibitem[{Shan and Wang(2008)}]{Shan2008}
Shan S, Wang G (2008) Reliable design space and complete single-loop
  reliability-based design optimization. Reliab Eng Sys Safety 93(8):1218--1230

\bibitem[{Song et~al.(2009)Song, Lu, and Qiao}]{Song2009}
Song S, Lu Z, Qiao H (2009) Subset simulation for structural reliability
  sensitivity analysis. Reliab Eng Sys Safety 94(2):658--665

\bibitem[{Sudret and {Der Kiureghian}(2002)}]{Sudret2002}
Sudret B, {Der Kiureghian} A (2002) Comparison of finite element reliability
  methods. Prob Eng Mech 17:337--348

\bibitem[{Taflanidis and Beck(2009{\natexlab{a}})}]{Taflanidis2009a}
Taflanidis A, Beck J (2009{\natexlab{a}}) Life-cycle cost optimal design of
  passive dissipative devices. Structural Safety 31(6):508--522

\bibitem[{Taflanidis and Beck(2009{\natexlab{b}})}]{Taflanidis2009}
Taflanidis A, Beck J (2009{\natexlab{b}}) Stochastic subset optimization for
  reliability optimization and sensitivity analysis in system design. Computers
  \& Structures 87(5-6):318--331

\bibitem[{Tu et~al.(1999)Tu, Choi, and Park}]{Tu1999}
Tu J, Choi K, Park Y (1999) A new study on reliability-based design
  optimization. J Mech Des 121:557--564

\bibitem[{Vazquez and Bect(2009)}]{Vazquez2009}
Vazquez E, Bect J (2009) {A sequential Bayesian algorithm to estimate a
  probability of failure}. In: 15th IFAC Symposium on System Identification,
  IFAC, Saint-Malo

\bibitem[{Waarts(2000)}]{Waarts2000}
Waarts PH (2000) {Structural reliability using finite element methods: an
  appraisal of {DARS}: {D}irectional {A}daptive {R}esponse {S}urface
  {S}ampling}. PhD thesis, Technical University of Delft, {The Netherlands}

\bibitem[{Welch et~al.(1992)Welch, Buck, Sacks, Wynn, Mitchell, and
  Morris}]{Welch1992}
Welch W, Buck R, Sacks J, Wynn H, Mitchell T, Morris M (1992) Screening,
  predicting, and computer experiments. Technometrics 34(1):15--25

\bibitem[{Youn and Choi(2004)}]{Youn2004a}
Youn B, Choi K (2004) Selecting probabilistic approaches for reliability-based
  design optimization. AIAA Journal 42:124--131

\end{thebibliography}

\end{document}